\documentclass[12pt,draftcls,onecolumn]{IEEEtran}
\usepackage{graphicx}
\usepackage{bm}
\usepackage{url}
\usepackage{epstopdf}
\usepackage{cite}
\usepackage{amsfonts,amssymb,amsmath}
\usepackage[usenames, dvipsnames]{color}
\usepackage{flushend}
\usepackage[]{algorithm2e}
\allowdisplaybreaks
\IEEEoverridecommandlockouts

\newtheorem{remark}{Remark}
\newtheorem{theorem}{Theorem}

\def\bpi{b_i}

\newcommand{\RN}[1]{%
  \textup{\uppercase\expandafter{\romannumeral#1}}%
}

\def\plotwidth{3.9in}

\begin{document}

\title{Broadcast Coded Modulation: Multilevel and Bit-interleaved Construction}

\author{Ahmed Abotabl, {\em Student Member, IEEE}, and Aria Nosratinia, {\em Fellow, IEEE}
\thanks{This work was supported in part by the grant CIF1219065 and ECCS1546969 from the National Science Foundation.}
\thanks{The authors are with the Department of Electrical Engineering, University of Texas at
    Dallas, Richardson, TX 75083-0688 USA, E-mail:
    ahmed.abotabl@utdallas.edu;aria@utdallas.edu.
}}

\maketitle

\vspace{-0.5in}
\begin{abstract}
The capacity of the AWGN broadcast channel is achieved by superposition coding, but superposition of individual coded modulations expands the modulation alphabet and distorts its configuration. Coded modulation over a broadcast channel subject to a specific {\em channel-input} modulation constraint remains an important open problem. Some progress has been made in the related area of unequal-error protection modulations which can be considered single-user broadcast transmission, but it does not approach all points on the boundary of the capacity region. This paper studies broadcast coded modulation using multilevel coding (MLC) subject to a specific channel input constellation. The conditions under which multilevel codes can achieve the constellation-constrained capacity of the AWGN broadcast channel are derived. For any given constellation, we propose a pragmatic multilevel design technique with near-constellation-constrained-capacity performance where the coupling of the superposition inner and outer codes are localized to each bit-level. It is shown that this can be further relaxed to a code coupling on only one bit level, with little or no penalty under natural labeling. The rate allocation problem between the bit levels of the two users is studied and a pragmatic method is proposed, again with near-capacity performance. In further pursuit of lower complexity, a hybrid MLC-BICM is proposed, whose performance is shown to be very close to the boundary of the constellation-constrained capacity region. Simulation results show that good point-to-point LDPC codes produce excellent performance in the proposed coded modulation framework.
\end{abstract}

\section{Introduction}

The capacity of the AWGN broadcast channel is achieved via superposition coding~\cite{Cover_BC,ELGamal:book}, but superposition of coded modulations is in general a modulation with much bigger size, and growth in the cardinality of constellation has practical costs that get progressively worse with more users. Quite aside from questions of cardinality, a superposition of coded modulations yields an irregular modulation constellation, with associated inconvenience and computational issues for the calculation of LLRs in hardware or firmware. Finally, the configuration of a superposition of constellations does not stay fixed throughout the rate region, in particular the peak-to-average power ratio (PAPR)~\cite{McCune15}, an important parameter for the efficiency of power amplifiers, becomes a variable quantity thus creating complications in the design of the transmitter.

Thus, broadcast coded modulation subject to a pre-determined transmit constellation is an important problem. 
Coded modulation in the  point-to-point channel has a long history and has been studied in great detail~\cite{Ungerboeck,Ungerboeck2,Forney:IT98}, but in the multi-node scenario, coded modulation introduces new and interesting phenomena and despite some progress, the design of capacity-approaching coded modulation  for the broadcast channel under a {\em channel-input} constellation constraint has remained an essentially open problem. An outline of related work is as follows. Taubin~\cite{BICM_BC} proposed the transmission of a weighted sum of two independent bit interleaved coded modulations and Sun {\em et al.}~\cite{Wesel} proposed superposition Turbo TCM for the broadcast channel. Neither of these  strategies obey a channel-input constellation constraint. A related area is the so-called single-user broadcasting~\cite{Shamai_SUBC}, where two streams are transmitted into a single-user channel with unequal-error protection (UEP). Earlier work in this area include Ramchandran {\em et al.}~\cite{Ramchandran}, on UEP modulation, however, the focus of their work is on providing variable error rates and not on capacity-approaching performance (see~\cite[Table II]{Ramchandran}).

\begin{figure*}
\centering
\includegraphics [scale=1]{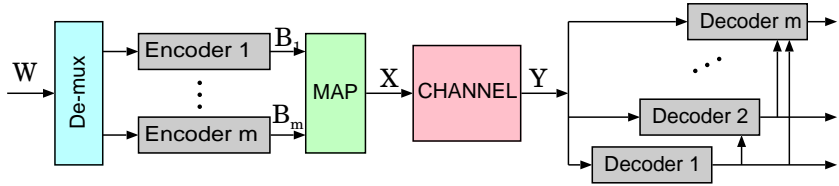}
\caption{MLC and MSD in point to point channel.}
\label{MLCSISO}
\end{figure*}

This paper addresses the design of multilevel coding (MLC) for the two-user AWGN broadcast channel under fixed constellation (in size and shape) at the channel input. In addition, a relative of MLC,  the bit-interleaved coded modulation (BICM)~\cite{BICM} is employed for efficient implementation. 
For a two-user broadcast channel, we refer to the superposition code component for the weak user (experiencing lower signal-to-noise ratio) as the ``outer code'' and for the strong user as the ``inner code.'' We show that for the inner code to be decomposable to multilevel code, necessary and sufficient conditions are essentially similar to the point-to-point scenario. We then show the optimality conditions for a multilevel decomposition of the outer code, and finally we highlight the optimality conditions for the (simultaneous) multilevel decomposition of the inner and outer codes. We show via numerical results that separating the two users' signals into distinct levels is in general insufficient to approach capacity. As mentioned earlier, this is the approach most commonly taken by the unequal error protection modulation schemes. Since mixing of the two users' signals is unavoidable, this paper proposes a simple level-wise concatenation of user's codewords that closely approaches the capacity limit. The mixing of the two users' data can be limited to only one of the levels. We also propose a hybrid MLC-BICM that further simplifies the design, yet has excellent performance. Finally, we show that good point-to-point codes can be used as component codes for the multilevel encoder with excellent performance. For more than two-users, there will be more than two layers of encoders. Each layer encodes the information of a different receiver. Necessary and sufficient conditions for the decomposition of each layer into multilevel decomposition is a straight forward extensions of the results of this paper. However, the design of bit-wise combining of more than two messages and the rate allocation per user at each level is not considered in this paper.

A brief background survey on multilevel coding is as follows: Multilevel coding was proposed by Imai and Hirakawa in~\cite{Imai}. More details about the performance and the design of MLC can be found in~\cite{Huber,huber1994capacities,MLC_Design}. Duan {\em et al.}~~\cite{Urbanke_MLC} showed that MLC with linear mapping does not require active shaping to achieve the capacity. The MLC error exponent was analyzed by Ingber and Feder~\cite{Feder_MLC}. MLC was extended to the MIMO transmission~\cite{MLC_MIMO}, was used for diversity coding~\cite{MLC_Diversity_coding,Sym_MDC,Asym_MDC,new_MDC} and in data storage~\cite{AbdelGhaffar}. Much less is known about MLC in the context of multi-node networks. A notable exception is~\cite{Narayanan} which used MLC in the context of compute and forward. But in general the optimality and efficient design of MLC for a variety of channels, including in particular the broadcast channel, has been for the most part an open problem. A primitive version of multilevel superposition was proposed by the present authors in~\cite{Attia:ISIT14}.

\section{Preliminaries}

 Multilevel coding is a coded modulation in which each input to the constellation mapper is driven by an independent encoder. When the encoders are binary and the constellation is $q$-ary, there are $m=\log_2(q)$ encoders. At each instant a bit is collected from the output of the encoders to form the vector $[B_1,\ldots,B_m]$ which will be mapped to point $X$ in the constellation (See Fig.~\ref{MLCSISO}) where throughout the paper, we use upper case letters to denote the random variables and lower case letters to denote the realization of the random variable.

Since the modulation mapping is bijective, the data processing inequality is fulfilled with equality:
\begin{equation}
\label{p2p-MLC}
I(X;Y)=I(B_1, B_2,\dots ,B_{m};Y)=\sum_{i=1}^{m}I(B_i;Y|B^{i-1})
\end{equation}
where $Y$ is the received signal and we denote the partial vectors $B^{i-1}=[B_1, B_2, \dots, B_{i-1}]$ in a manner similar to~\cite{Cover06:book}. It was shown by Ingber and Feder~\cite{Feder_MLC} that multilevel coding achieves the constellation constrained capacity if and only if the input optimal distribution can be expressed as the multiplication of the marginal distribution of each of the bits driving each level $P_X^*(x)=\prod_{i=1}^mP_{B_i}(b_{i})$ where $P^*$ denotes the optimal distribution. The right hand side in \eqref{p2p-MLC} justifies multistage decoding. Multistage decoding is implemented by decoding $B_i$ conditioned on $[B_{i-1},\ldots,B_1]$. Therefore, the rate of level-$i$ should always satisfy 
\begin{equation}
R_i\leq I(B_i;Y|B^{i-1})
\end{equation}
in order to achieve a vanishing error probability where $R_i$ is the rate of encoder $i$. Subject to choosing the appropriate rates, the constellation constrained capacity can be achieved, which itself, subject to appropriate choice of constellation, can approach the channel capacity.

In this paper we consider multilevel coding in the context of the degraded Gaussian broadcast channel, in particular using superposition coding~\cite{Cover_BC}.

Throughout the paper, the SNR of a point-to-point AWGN channel is denoted by $\rho$ and the SNR of the weak and the strong receivers of the AWGN broadcast channel are denoted by $\rho_1$ and $\rho_2$ respectively. Also, the noise variance at the weak and the strong receivers are denoted by $\sigma_1^2$ and $\sigma_2^2$.

\section{Analysis of Multilevel Superposition Coded Modulation}

\subsection{Multilevel Inner Code}

\begin{figure*}
\centering
\includegraphics[scale=1.1]{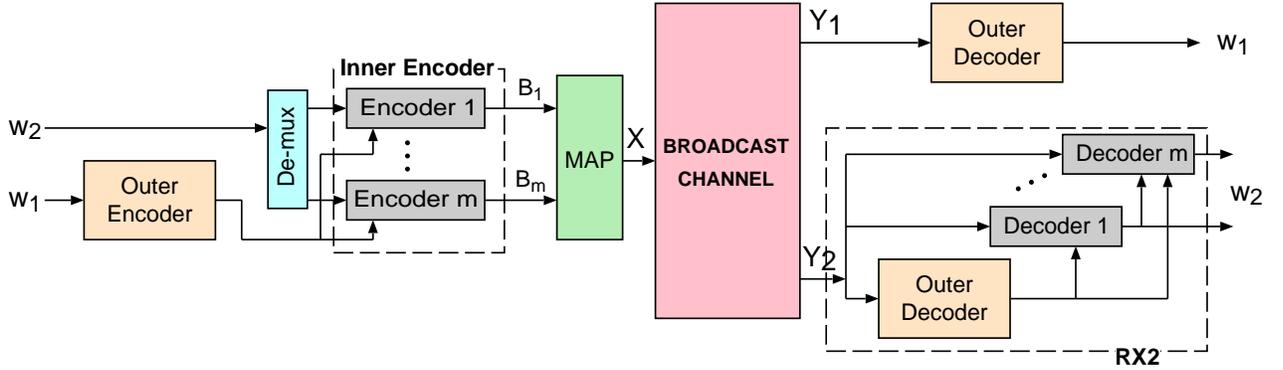}
\caption{Broadcast channel with MLC for the inner code where every encoder codebook is a realization of random generation}
\label{fig:BC1}
\end{figure*}

We begin by investigating multilevel decomposition of the inner code (see Fig.~\ref{fig:BC1}). The message $w_1$ is encoded with the outer code which is generated according to a distribution $p_U(u)$ to give the cloud centers of the superposition code (the codewords that will be decoded at both receivers). The message $w_2$ is split into $m$ sub-messages. Sub-message $i$ is encoded with inner code at level $i$ that is generated according to a distribution $P_{B_i|U}(b_i|u)$. The inner code obeys an alphabet constraint on $X$ as well as a multilevel coding constraint on the individual bits representing $X$, while the outer code in this case is unconstrained. The question is: under what conditions can such a decomposition meet the constellation constrained capacity?

The channel input $X$ is constrained to a specific constellation via a one-to-one function $f: [B_1,\cdots ,B_m] \rightarrow X$ whose domain is a vector of coded bits $[B_1,\cdots ,B_m]$. The achievable rate region of the broadcast channel subject to multilevel coding constraint on the inner code can be characterized by the following collection of weighted sum rates:
\begin{align}
\label{sum_rate_MLC}
R=\max_{\prod_{i=1}^mP_{B_i|U}(\bpi|u)P_U(u)}\{ \theta I([B_1,\cdots ,B_m];Y_2|U)+(1-\theta) I(U;Y_1)\}
\end{align}
where $\theta \in[0, 1]$  is a parameter indicating the point achieved on the boundary of the rate region.

The modulation-constrained sum rate for the two-user degraded broadcast channel {\em without} any multilevel coding constraints is given by
\begin{align}
\label{sum_rate_general}
R=\max_{P_{B_1,\cdots,B_m|U}(b_1,\cdots ,b_m|u)P_U(u)} \{ \theta I([B_1,\cdots ,B_m];Y_2|U)+(1-\theta) I(U;Y_1)\}
\end{align}
where the difference of  \eqref{sum_rate_MLC} and \eqref{sum_rate_general} is that the former is optimized over a product conditional distribution for $B_1,\cdots,B_m$, whereas the latter is optimized over a general distribution. If the two sum-rate expressions are identical for all values of $\theta$, it follows that the capacity regions must be identical.

\begin{theorem}
A multilevel inner code achieves the constellation constrained capacity of the degraded broadcast channel if the capacity-achieving distributions on the individual bits of the modulation are conditionally independent, i.e.,
\begin{equation}
P_{B_1,\cdots,B_m|U}^*(b_1,\cdots,b_m|u)=\prod_{i=1}^mP_{B_i|U}^*(b_i|u)
\end{equation}
\end{theorem}

This optimality result is the counterpart of the point-to-point optimality result of Ingber and Feder~\cite{Feder_MLC}. The individual rates can be calculated using the usual peeling decoder for the strong user. When the outer decoder is implemented via multistage decoding, the achievable rates are:
\begin{align}
R_1&\leq I(U;Y_1)\\
R_2&\leq I(X;Y_2|U)=\sum_{i=1}^mI(B_i;Y_2|U,B^{i-1})
\end{align}
It follows that multistage decoding of the inner code is possible when
\begin{equation}
R_{2i}\leq I(B_i;Y_2|U,B^{i-1})
\end{equation}
where $R_{2i}$ is the rate of the inner encoder at level $i$.

\subsection{Multilevel Outer Code}
\label{Sec:Multilevel-outer}

We now consider the case when the inner code is unconstrained, but the outer code is a multilevel code. The outer code represents the cloud centers and is generated by the auxiliary random variable $U$, whose cardinality is enough to be bounded by the cardinality of $X$ for optimality. The question is: when can the outer code be decomposed into {\em independently encoded} levels?

\begin{figure*}
\centering
\includegraphics[scale=1.1]{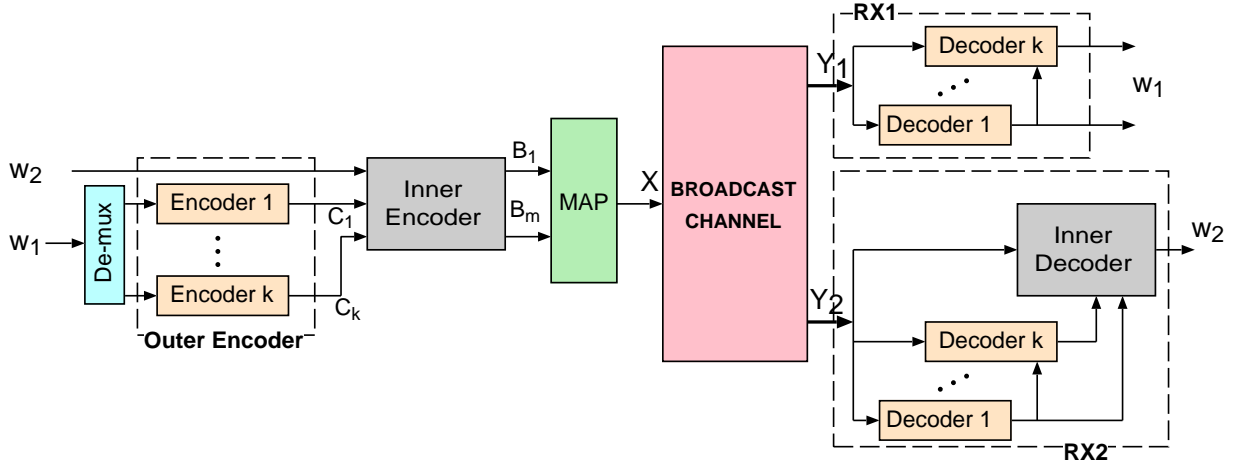}
\caption{Broadcast channel with multilevel coding for the outer code where every encoder codebook is a realization of a random generation}
\label{fig:BC2}
\end{figure*}

We now argue that it is always possible to produce a multilevel decomposition of the outer code with arbitrarily small loss, as long as it is permissible to increase the number of coding levels.

Consider a set of binary variables $C_1,\ldots,C_k$ representing the levels of the inner code, drawn independently according to Bernoulli-$\frac{1}{2}$. We now aim to find a mapping $g: [C_1, \cdots, C_k] \rightarrow U$ such that $p_U(u)$ approximates the capacity-optimizing distribution $p^*_U(u)$. Since each realization of $C^k$ has probability $2^{-k}$, the design of $g(\cdot)$ consists of crafting a many-to-one mapping from the bit vector to $U$ so that 
\[
2^{-k}  \big|\{ [c_1, \cdots, c_k] \; : \; g(c_1, \cdots, c_k)=u_i\}\big| \approx P^*_{U}(u_i)
\]
where $|\cdot|$ stands for the cardinality of the set it contains, and $P_{U^*}(u)$ is the optimal distribution of $P_{U}(u)$. It is not difficult to see that one is guaranteed to get to within $2^{-k}$ of approximating each $p_U(u)$.

The individual rates  are therefore:
\begin{align}
R_1&\leq I(U;Y_1)=\sum_{i=1}^k I(C_i;Y_1|C^{i-1})\\
R_2&\leq I(X;Y_2|C^k)
\end{align}
where $U=g([C_1, \cdots, C_k])$. Multistage decoding of the outer code at both receivers is subject to the following individual rate constraints
\begin{equation}
\label{rate_const1}
R_{1i}\leq I(C_i;Y_1|C^{i-1})
\end{equation}
where $R_{1i}$ is the rate of the encoder in level $i$ of the outer encoder. Intuitively, if the weak receiver can do multistage decoding at a certain set of rates, so can the strong receiver at the same set of rates, because the strong receiver is less noisy. Formal derivation of this fact is straightforward and is relegated to appendix~\ref{Appendix:OuterCodeDecomposition}.

\begin{figure*}
\centering
\includegraphics[scale=1]{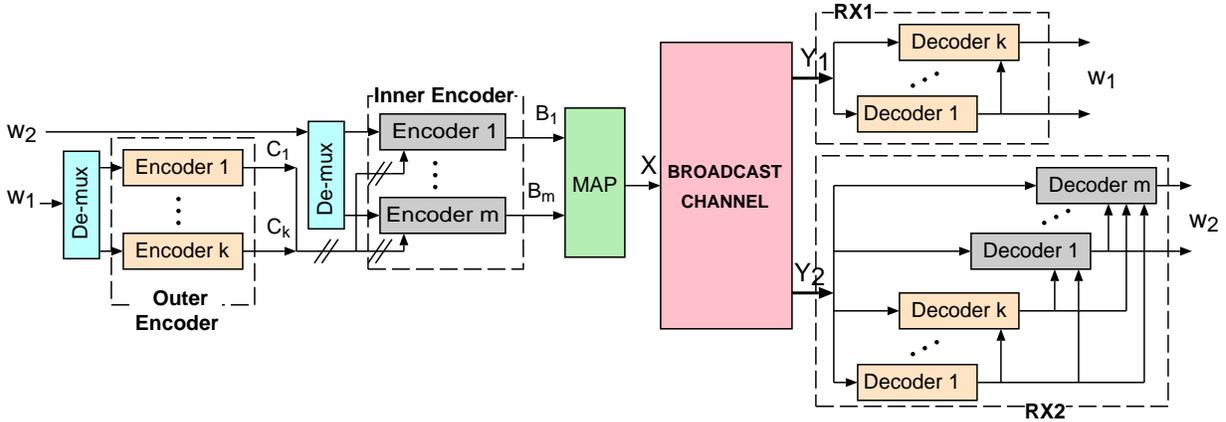}
\caption{Broadcast channel with full multilevel superposition coding where every encoder codebook is a realization of random generation}
\label{fig:BC3}
\end{figure*}

\subsection{Full Multilevel Superposition Coding}

We now consider the case when the outer and the inner codes are decomposed to multilevel construction (see Fig.~\ref{fig:BC3}). Each encoder in the inner code depends on its message and the output of all the encoders of the outer code. The maximum achievable sum rate is given by
\begin{equation}
R=\max_{\prod_{i=1}^mP_{B_i|C^k}(b_i|c^k)P_{C_i}(c_i)} \theta I([B_1, \dots, B_m];Y_2|U)+(1-\theta) I([C_1, \dots, C_k];Y_1)
\end{equation}
Denote the  optimal distribution under the channel input constraint $X=f(B_1,\cdots,B_m)$ with $P^*_{X|U}(x|u)P_U(u)=P^*_{B_1,\cdots,B_m|U}(b_1,\cdots,b_m|u)P_U(u)$. A necessary and sufficient condition for the constellation-constrained optimality of a multilevel decomposition is that there exists a (potentially many-to-one) function $g(\cdot)$ so that for every $u$,
\begin{equation}
\label{SC_MLC}
P^*_{B_1,\cdots,B_m|U}(b_1,\cdots,b_m|u)P_U(u)=\sum_{g(c^k)=u}\prod_{i=1}^mP_{B_i|C^k}(b_i|c^k)\prod_{j=1}^kP_{C_j}(c_j)
\end{equation}
This means that the capacity achieving distribution on the coded bits $B_1,\cdots,B_m$ can be constructed by, firstly, cloud centers generated via independent binary variables $C_1,\ldots,C_k$ together with a mapping $g:C^k\rightarrow U$, and secondly coded bits $B_1,\ldots,B_m$ that are independent {\em conditioned on} $C_1,\ldots,C_k$.  Using arguments similar to the ones in Section~\ref{Sec:Multilevel-outer} and Appendix~\ref{Appendix:OuterCodeDecomposition}, one can show that the conditions on the outer code can be satisfied to any required degree of approximation via increasing $k$, the number of the levels of the outer code.

Under this condition, the individual rates are:
\begin{align}
R_1&\leq I(U;Y_1)=\sum_{i=1}^kI(C_i;Y_1|C^{i-1})\\
R_2&\leq I(X;Y_2|C^k)=\sum_{j=1}^mI(B_j;Y_2|B^{j-1},C^k)
\end{align}
Multistage decoding of the outer and inner codes at both receivers is subject to the following individual rate constraints
\begin{align}
R_{1i}&\leq I(C_i;Y_1|C^{i-1}) & 1\le i \le k\\
R_{2j}&\leq I(B_j;Y_2|B^{j-1},C^k) & 1 \le j \le m
\end{align}

\section{Design of Multilevel Superposition Coded Modulation}

The results of the previous section show the conditions under which broadcast capacity can be achieved by multilevel coding. The remainder of this paper shows that even in the absence of optimality conditions, MLC can still achieve rates very close to the boundary of the capacity region. This section produces a design methodology for multilevel broadcast coded modulation via a simple coding framework that greatly facilitates the design process and yet induces little or no performance penalty (allows near-optimal performance). Subsequently, we solve the problem of rate allocation between the users and layers of the multilevel code in the context of the proposed framework, thus completing the design process.

\subsection{Bit-additive Superposition coding}
\label{sec:Bit-wise}

In the multilevel decomposition considered so far, each of the inner encoder levels depends on the code vector produced by {\em all} the outer encoders. The cross dependency of multiple codes is difficult to implement in practice, therefore it is natural to seek encoding methods whose levels are decoupled from each other {\em for both users}, especially considering that the notion of decoupling of levels is at the heart of motivation for the point-to-point multilevel codes~\cite{Imai}. This means that level-$i$ encoder of the inner code reads only the output of level-$i$ outer encoder, which leads to a {\em bit-wise superposition}. This can be optimal only if, in addition to the condition~\eqref{SC_MLC}, we also have:
\begin{equation}
P_{B_i|C^k}(b_i|c^k)=P_{B_i|C_i}(b_i|c_i) \qquad \forall i
\end{equation}
For most modulations used commonly in practice, this condition cannot be met precisely. Nevertheless, it is possible to achieve performance very close to capacity via an encoding method that decouples the bit levels from each other, and furthermore implements the superposition at each level by a simple binary additive operation. We call this simple multilevel superposition strategy the {\em bit-additive superposition}. We now proceed to describe this method and demonstrate its performance.

\begin{figure}
\centering
\includegraphics[scale=1]{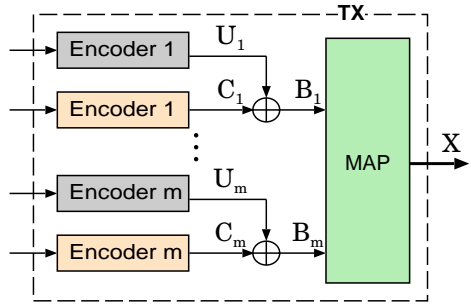}
\caption{XOR implementation of multilevel bit-wise superposition coding.}
\label{fig:BC_XOR}
\end{figure}

Fig~\ref{fig:BC_XOR} shows the outline of the proposed method. The outer codes are generated independently according to  Bernoulli-$\frac{1}{2}$ distribution, each with a prescribed rate $R_{1i}$, and are represented with variable $C_i$. The inner codes are represented by $U_i$, which are generated independently according to the distribution Bernoulli-$\alpha_{i}$ with $\alpha_i\in [0,0.5]$. Bit-additive superposition is achieved via $B_i = C_i \oplus U_i$ where $\oplus$ represents the binary XOR operation. When $\alpha_i=0$, we have $B_i=C_i$ so we have $R_{2i}=0$. When $\alpha_i=0.5$, $B_i$ is independent of $C_i$ and $R_{1i}=0$. This method of binary superposition is mentioned, among others, in~\cite[Chapter 5]{ELGamal:book} and~\cite{Shamai_XOR_superposition}.

The proposed bit-additive superposition can be implemented in the following manner: a binary linear code is chosen for each level of the outer code since linear codes have uniform distribution. For the encoders of the inner code, we need a code with distribution Bernoulli-$\alpha_i$. Such a code can be generated from a linear code which has a uniform distribution and set the bits at randomly chosen locations with zero. For example, if the required distribution is Bernoulli-$\alpha_i$, then the number of bits set to zero (regardless of their original value) should be
\begin{equation}
N=2(1/2-\alpha_i)n
\end{equation}
where $n$ is the block-length of the code.

\subsection{Performance of Bit-additive Superposition}

We now provide numerical examples for a wide variety of modulations to demonstrate the efficacy of the proposed bit-additive superposition. The general setup for these numerical studies is as follows.

\begin{figure}
\centering
\includegraphics[width=\plotwidth]{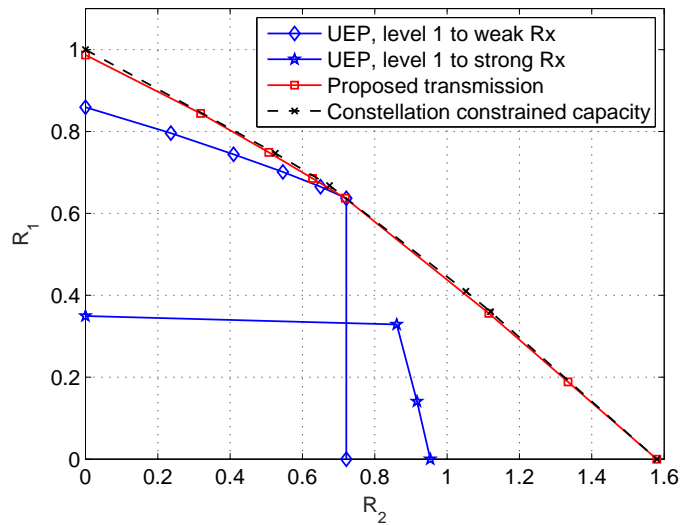}
\caption{Comparison of proposed technique with UEP-type modulation that assigns levels to distinct users under $4$-PAM, $\rho_1=5dB$, $\rho_2=10dB$. Proposed technique has negligible gap to constellation constrained capacity, while UEP-type modulations can be far from capacity.}
\label{fig:Sim1}
\end{figure}

\begin{figure}
\begin{minipage}{\columnwidth}
\centering
\includegraphics[width=\plotwidth]{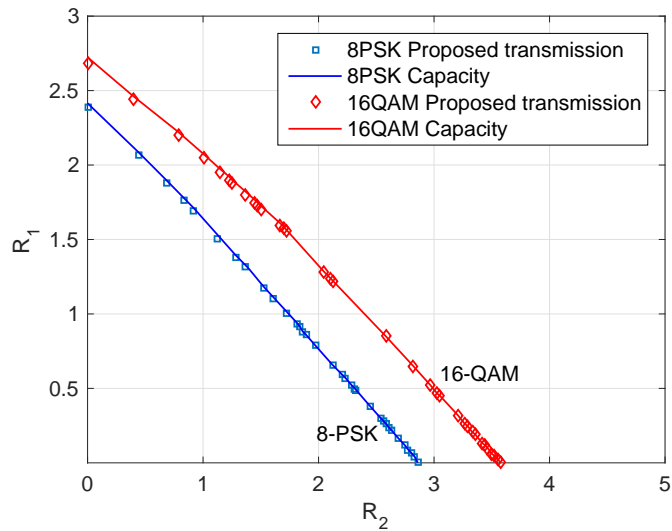}
\caption{Proposed MLC transmission rates for $8$-PSK and $16$-QAM where $\rho_1=8dB$ and $\rho_2=12dB$. The proposed MLC technique is indistinguishable from constellation constrained capacity in each case.}
\label{fig:2dim_const}
\end{minipage}
\end{figure}

The baseline for comparisons in each case is the constellation constrained capacity, which is calculated using the modified Blahut-Arimoto algorithm~\cite{Const_limit_cap}. In each case, the achievable rate region for the proposed bit-additive superposition is obtained in the following manner: For each level $i$, a  uniformly distributed codeword is generated for the weak receiver and a codeword with distribution Bernoulli-$\alpha_i$ for the strong receiver. The input to the mapper at level $i$ is the XOR between the weak receiver codeword at level $i$ and the strong receiver codeword at level $i$. Each value of the vector $[\alpha_1, \alpha_2, \dots, \alpha_m]$ gives a certain rate pair $(R_1,R_2)$. For every value of the vector $[\alpha_1, \alpha_2, \dots, \alpha_m]$, the mutual informations
\begin{align}\nonumber
&I(C_1, \dots, C_m;Y_1)\\\nonumber
&I(B_1, \dots, B_m;Y_2|C_1, \dots, C_m)
\end{align}
are calculated. These mutual informations give an achievable rate pair $R_1$ and $R_2$ respectively.

Numerical results show a very small gap between constellation constrained capacity and the proposed bit-additive superposition. In particular Figure~\ref{fig:Sim1} for the $4$-PAM constellation, and Fig.~\ref{fig:2dim_const} shows the performance of bit-additive superposition for $16$-QAM and $8$-PSK. Simulations show the same achievable rate region via Gray and natural mapping.

Fig.~\ref{fig:Sim1} also shows comparisons to a bit-allocation strategy often used by the Unequal-Error Protection (UEP) modulations~\cite{Ramchandran,UEP_MLC}, i.e., the higher-order bit levels are assigned to one data category and the lower-order bit levels to the other data category. 

Fig.~\ref{fig:Sim1} represents $4$-PAM modulation, and the UEP-type modulation curves represent the two possibilities of level-1 (respectively level-2) being assigned to weak (respectively strong) user, or vice versa.  In the former case, we see that this assignment meets the capacity outer bound only at one point, otherwise it can be far from capacity. 
Reversing the assignment of modulation index to the users results in even worse performance. 

It has been noted by~\cite{Lee-Fang,CM_sat_BC,Shu-Lin} that in the UEP approach one may allocate each modulation index to one message at a time, but then allow time sharing between all such strategies. Thus one may achieve the convex hull of all points on such individual rate assignments, as well as the single-user rates. This can provide a performance closer to capacity, but requires buffering with its associated additional delay.

\begin{remark}
For a fixed channel SNR and for a fixed rate pair, the larger the modulation size, the smaller is the gap-to-capacity for a static assignment of messages to modulation indices.
\end{remark}

\begin{figure}
\centering
\includegraphics[width=\plotwidth]{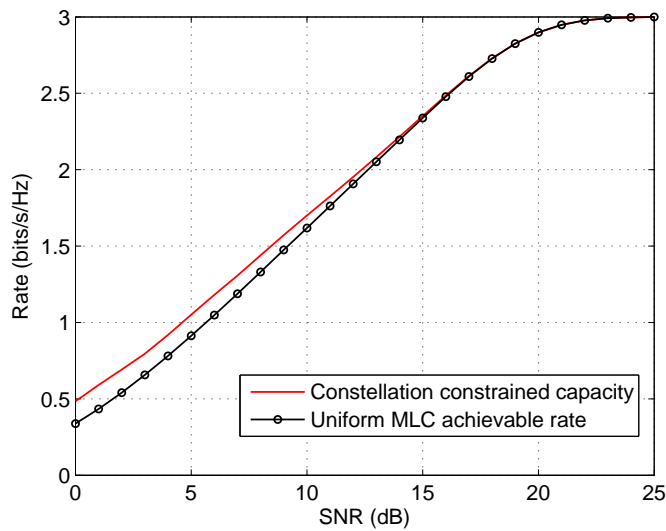}
\caption{The penalty for using multilevel {\em linear} coding (equi-probable zeros and ones) in a single-user channel under $8$-PAM with natural labeling}
\label{fig:MLC-gap}
\end{figure}

\begin{remark}
In Fig.~\ref{fig:2dim_const} and even more so in Fig.~\ref{fig:Sim1}, there is a very small gap between the modulation-constrained capacity and the multilevel coding rates, especially close to the vertical axis (when the weak user mostly occupies the channel). 
This can be clarified by looking at  the single-user optimality condition of multilevel coding~\cite{Feder_MLC}, finding that it is not met for PAM with natural labeling. For the single-user $8$-PAM modulation under natural labeling, Fig.~\ref{fig:MLC-gap} shows the relationship of constellation constrained capacity and MLC achievable rate. 
8-PAM experiences a MLC penalty that is more severe at low SNR,\footnote{In the point-to-point channel this penalty goes away if at lower SNRs one uses a lower order modulation. Using a higher order modulation {\em and} requiring that all modulation points be used with equal probability (linear component codes) produces the rate penalty. In the broadcast channel this small penalty is not as easily avoidable because the same modulation is used to transmit to both users.} therefore the slight separation of rate curves in Figs.~\ref{fig:Sim1}, and~\ref{fig:2dim_const} is explained especially at the point where the weak user occupies the channel.
\end{remark}

\subsection{A Pragmatic Rate Allocation Algorithm}

To achieve a desired broadcast rate pair  $(R_1,R_2)$ in the context of multilevel coding, it is necessary to identify the relevant codes at each layer, which begins by specifying the code rates $R_{1i}$, $R_{2i}$ for all levels $i$. In this subsection, we present a pragmatic solution to this problem that in addition to its modest computational requirement, serves to reveal interactions between the rate constraints at different bit levels as well as interesting connections to the familiar single-user MLC mutual information curves. It will be demonstrated via simulations that this pragmatic method operates very close to the capacity region for most familiar modulations and mappings. Subsequently, we will discuss the rare cases where this pragmatic method may lead to a slight departure from optimality, and propose a general (but not as computationally thrifty) algorithm for rate allocation in such cases.

We begin by casting the rate allocation problem in the form of the following optimization, where $\theta$ parametrizes the boundary of the broadcast rate region:
\begin{align*}
\max_{\Pi_i P_{B_i|C^k}(b_i|c^k)\Pi_j P_{C_j}(c_j)} & \theta \sum_i R_{1i}+ (1-\theta) \sum_j R_{2j}\\
\text{Subject to }
&R_{1i}\leq I(C_i;Y_1|C^{i-1}) \qquad 1\le i \le k\\
&R_{2j}\leq I(B_j;Y_2|B^{j-1},C^k) \quad 1 \le j \le m \\
&R_{1i} \ge 0 \qquad R_{2j} \ge 0   \quad \forall i,j
\end{align*}
We will come back to a version of this general rate allocation problem in the sequel, but for now we concentrate on bit-additive superposition, where the rate allocation problem reduces to the following:
\begin{align}
\max_{\Pi_i P_{U_i}(u_i)P_{C_i}(c_i)} &\sum_i\theta R_{1i}+ (1-\theta) R_{2i}\\
\text{Subject to }
&R_{1i}\leq I(C_i;Y_1|C^{i-1}) \qquad 1\le i \le m \label{eq:ReliableDecoding1}\\
&R_{2i}\leq I(U_i;Y_2|U^{i-1},C^k) \quad 1 \le i \le m \label{eq:ReliableDecoding2}\\
&R_{1i} \ge 0 \qquad R_{2i} \ge 0 \nonumber
\end{align}
The key difference is that the maximization is now over independent distributions, therefore the utility function can now be decomposed into the sum of $m$ non-negative level-wise utility functions. 

\begin{figure}
\centering
\includegraphics[width=\plotwidth]{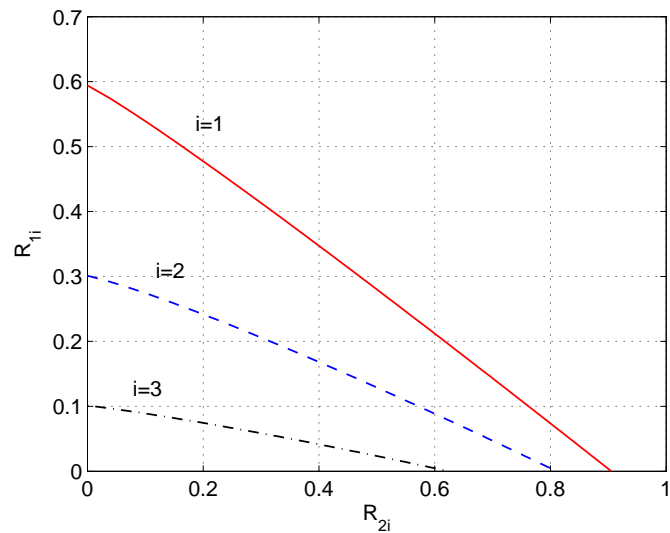}
\caption{Rate constraints for the levels of $8$-PAM constellation assuming natural labeling and decoding order from MSB to the LSB with $\rho_1=5$dB and $\rho_2=10$dB.}
\label{fig:8PAM-rate-constraints}
\end{figure}

Having arrived at a simplified utility function, we now concentrate on the constraints by highlighting the shape of the feasible rate regions at each individual level, which can be thought of as cross sections of the overall feasible rate region. For insight, we look into the specific example of $8$-PAM with natural labeling, where the level-wise rate constraints are shown\footnote{For each $i$, we have set the rates in other levels $j\neq i$ so that $R_{1j}=0$.} in Figure~\ref{fig:8PAM-rate-constraints}. 

\begin{figure}
\centering
\includegraphics[width=\plotwidth]{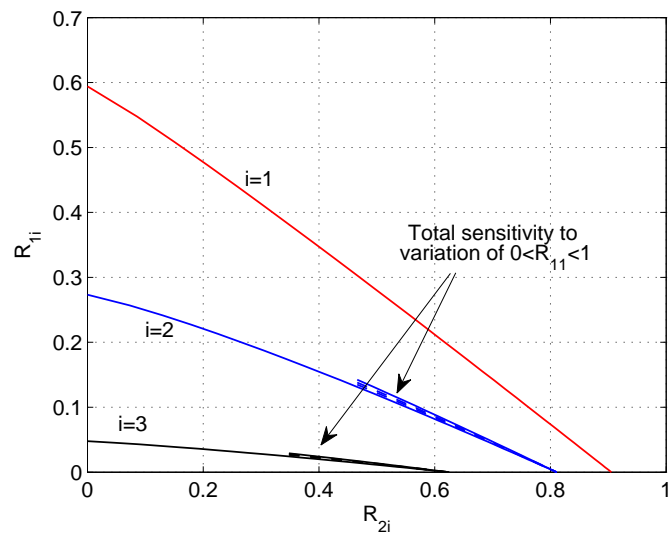}
\caption{Sensitivity of each level's constraint to rates of other levels}
\label{fig:constraint-sensitivity}
\end{figure}

\begin{figure}
\centering
\includegraphics[width=\plotwidth]{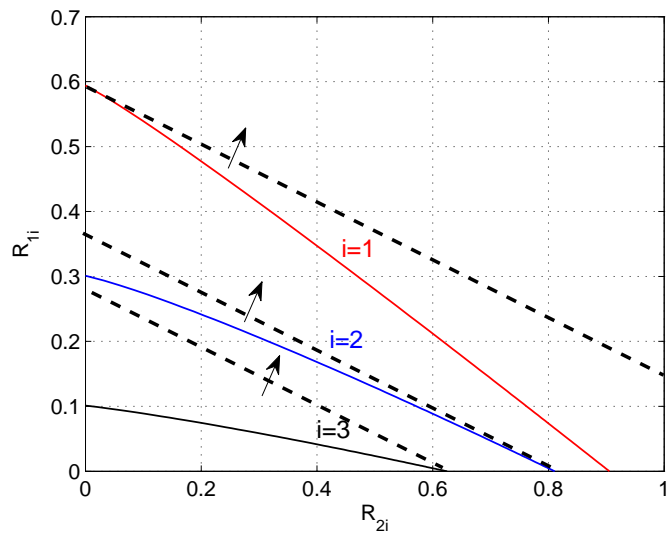}
\caption{Rate allocation via optimization at each level}
\label{fig:8PAM-optimization}
\end{figure}

The first interesting feature of the bit-level constraints is that, under most bit mappings including natural and Gray mapping, the binary rate constraint at each level  is largely insensitive to the parameters pertaining to other levels. For example, please see Figure~\ref{fig:constraint-sensitivity}, where in an $8$-PAM multilevel coded modulation, the sensitivity of the rate constraints in levels $2, 3$ at the set point $R_{22}=R_{32}=0$ is demonstrated subject to a complete sweep of the rate pair $R_{11},R_{12}$. From this observation rises a pragmatic assumption: that at optimality, one may assume that the constraints at different levels are approximately independent.\footnote{This approximation has been verified for all natural and Gray labeling for a variety of PAM, PSK, and QAM type modulations. There exist some irregular labeling for which this assumption fails. That case will be discussed separately in the sequel.} This approximation leads to a complete decomposition of the optimization into level-wise optimizations whose only coupling is through the parameter $\theta$, namely, for each $i=1, \ldots,m$,
\begin{align}
\max_{P_{U_i}(u_i)P_{C_i}(c_i)} &\theta R_{1i}+(1-\theta) R_{2i}\\
\text{Subject to }
&g_i(R_{1i},R_{2i}) 
\le 0\\
&R_{1i}\ge 0 \qquad R_{2i}\ge 0
\end{align}
where $g_i(\cdot,\cdot)$ is the rate constraint at each level whose dependence explicitly on $R_{1i},R_{2i}$ and omission of other variables is meant to highlight the approximate independence of the constraints at each level. Solving a typical rate allocation problem in the aforementioned example involves pushing a line with a slope determined by $\theta$ outward on the three levels mentioned above. An example is shown in Figure~\ref{fig:8PAM-optimization}, where the individual rate constraints for the three levels are shown in solid lines and the parallel dotted lines represent, for a fixed $\theta$, the lines $\theta R_{1i}+(1-\theta) R_{2i}=\alpha_i$, and the maximization of $\alpha_i$ corresponds to the movement of the dotted lines as shown by arrows.

The result of this rate allocation is that Level 1 is dedicated to User 1, and levels 2 and 3 are dedicated to User 2. Note that the rate constraint curves were calculated under the operating regime that all three levels are assigned to User 2. To take into account the (small) sensitivity of the individual rate regions to the operating point of other levels, one may update the three rate curves once more and verify that optimality conditions remain satisfied at the proposed optimal point. The update may slightly adjust the intercept points.

We now consider a second empirical property of level-wise  binary rate regions: that they are very nearly affine. This feature has been experimentally observed across modulations, bit level mappings, and various channel SNRs. The outcome of this second observation is that near optimal rate allocation can be achieved while allocating all the bits in each level to either one or the other user. This produces $2^m$ rate pairs that are close to the boundary of the rate region. Rate pairs in between can be achieved by dividing the rate in one of the levels (whose achievable rate slope is closest to $\theta R_{1} + (1-\theta) R_2$ between the two users.

\begin{figure}
\centering
\includegraphics[width=\plotwidth]{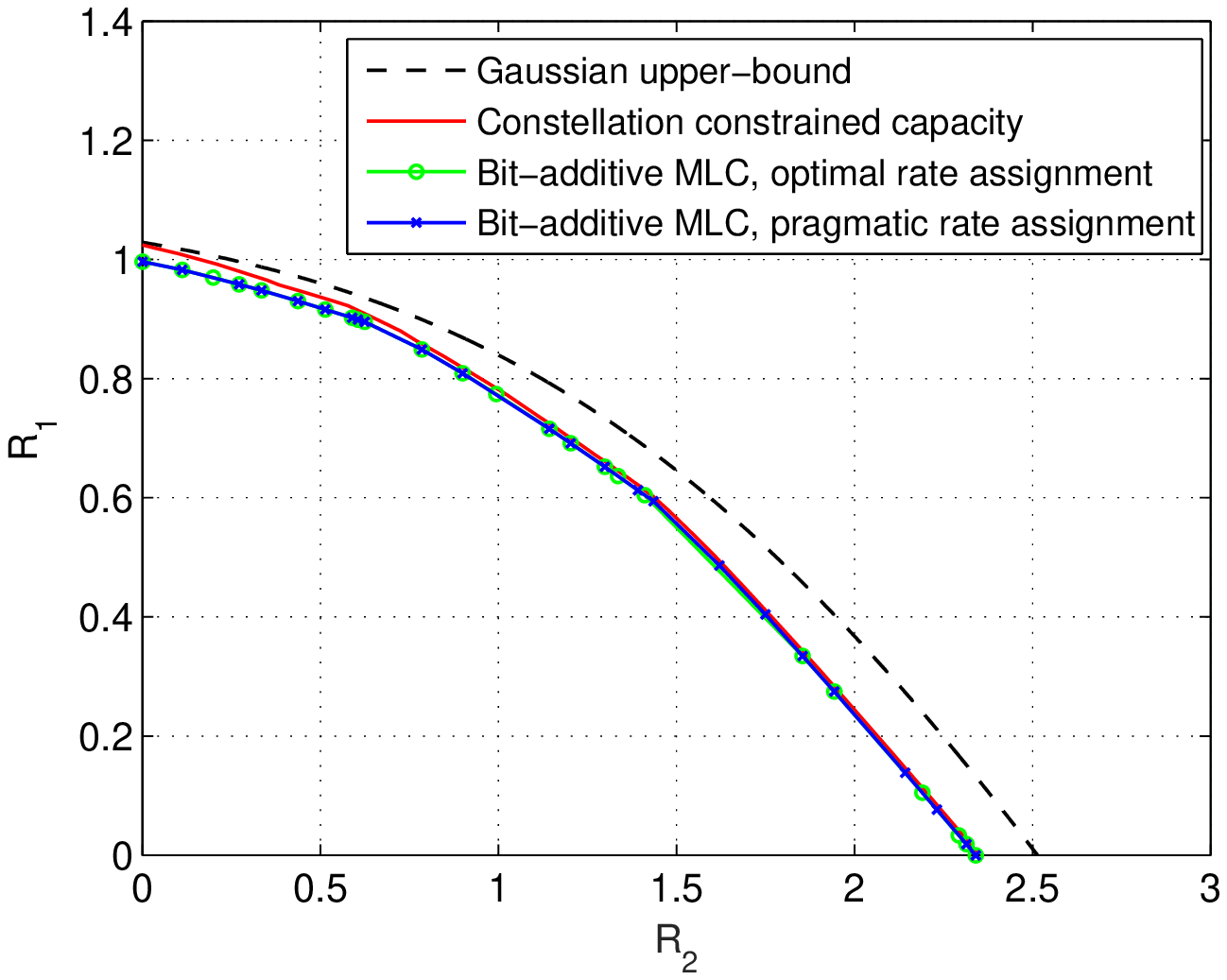}
\caption{MLC rate region for $8$-PAM, $\rho_1=5$dB, $\rho_2=15$dB.}
\label{fig:rate_assignment}
\centering
\includegraphics[scale=1]{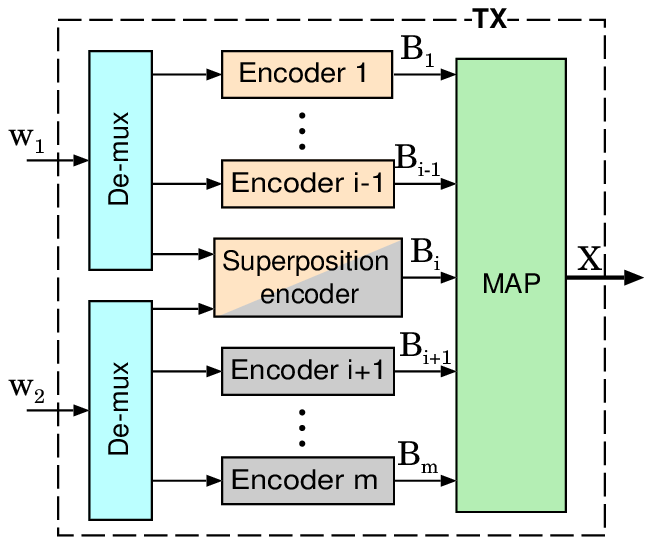}
\caption{Multilevel superposition with pragmatic rate allocation.}
\label{fig:MLC_BC_Efficient}
\end{figure}

This approach yields results that are practically indistinguishable from optimal rate allocation, with very few exceptions that are discussed in the next subsection. The performance of this method is illustrated, for the case of a $8$-PAM modulation with natural mapping, in Fig.~\ref{fig:rate_assignment}. In this figure, the normalized SNR of the two users are respectively $5$dB and $15$dB. The dotted line shows the Gaussian capacity without a modulation constraint. The red curve shows the modulation-constrained capacity that has been calculated via a variation of the Blahut-Arimoto algorithm. The achievable rate of the bit-additive multilevel coding is shown with the green plot, which is obtained by a full-search optimization for rate-allocation, potentially yielding a solution where each user's data is transmitted at all levels. The result of pragmatic rate allocation is shown with the blue plot, which is indistinguishable from the fully optimal rate allocation.

As noted earlier, the pragmatic rate allocation will result in a solution where most of the layers are allocated to one user or another, and potentially one level sees the data of both users. This will results in a solution that is shown in Fig.~\ref{fig:MLC_BC_Efficient}.

\begin{figure*}
\begin{minipage}{0.32\textwidth}
  \centering
  {\small \hspace{.3in} $4$-PAM }
\includegraphics[width=1.12\hsize]{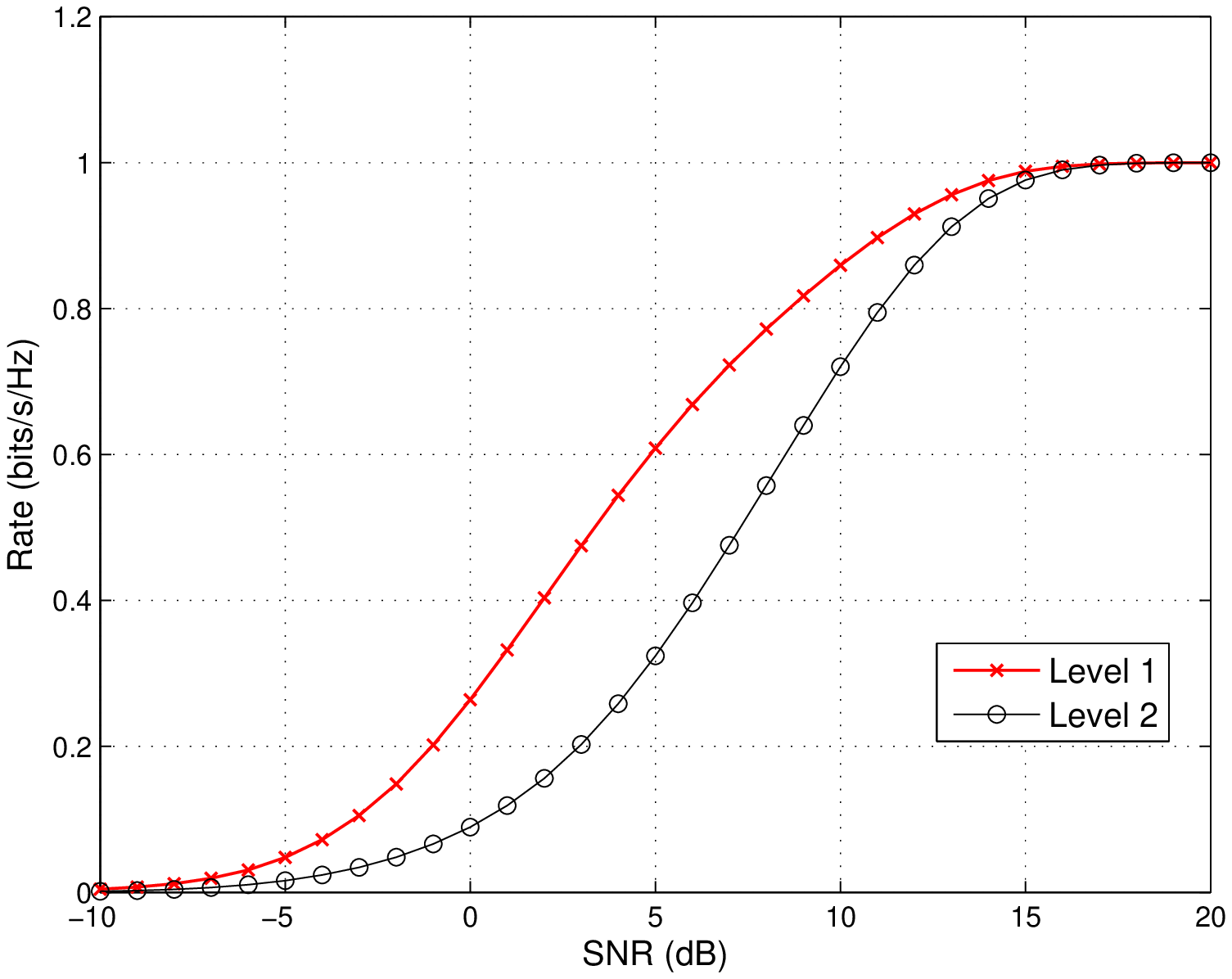}

\end{minipage}
\begin{minipage}{0.32\textwidth}
  \centering
	{\small \hspace{.3in} $8$-PAM }
  \includegraphics[width=1.12\hsize]{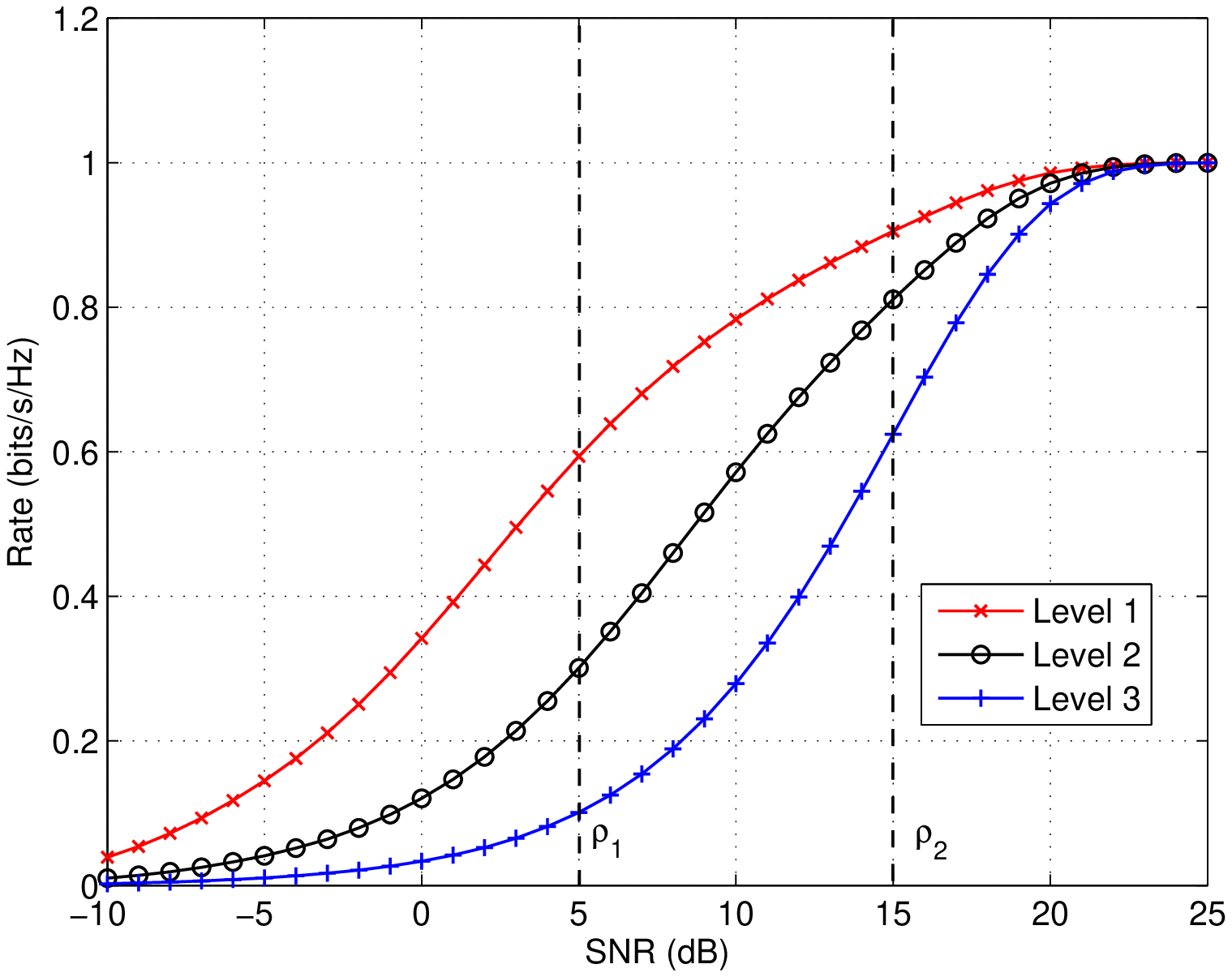}
\end{minipage}
\begin{minipage}{0.32\textwidth}
  \centering
	{\small \hspace{.3in} $16$-PAM }
  \includegraphics[width=1.12\hsize]{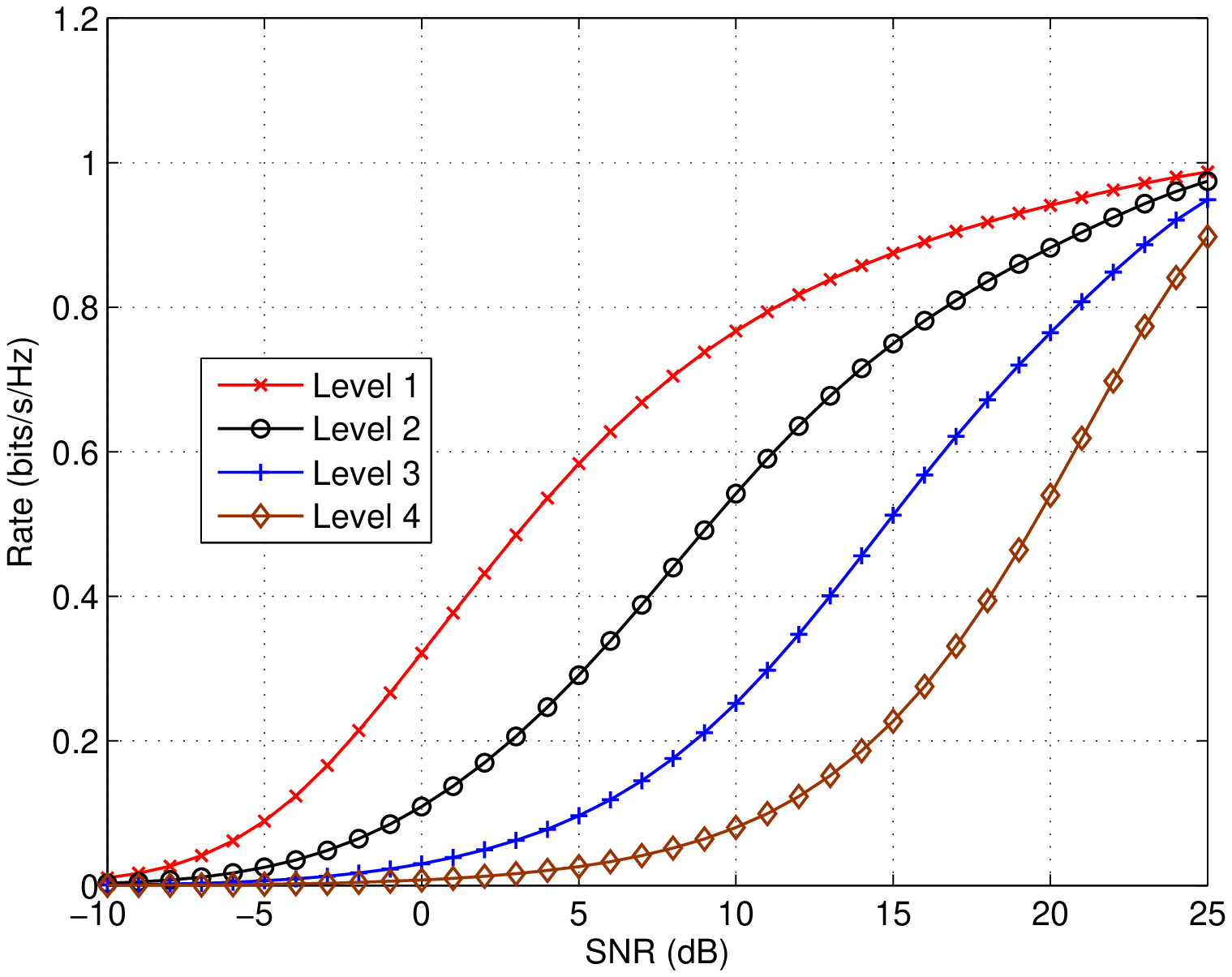}
\end{minipage}\\[.2in]
\begin{minipage}{0.32\textwidth}
  \centering
	{\small \hspace{.3in} $4$-PSK }
  \includegraphics[width=1.12\hsize]{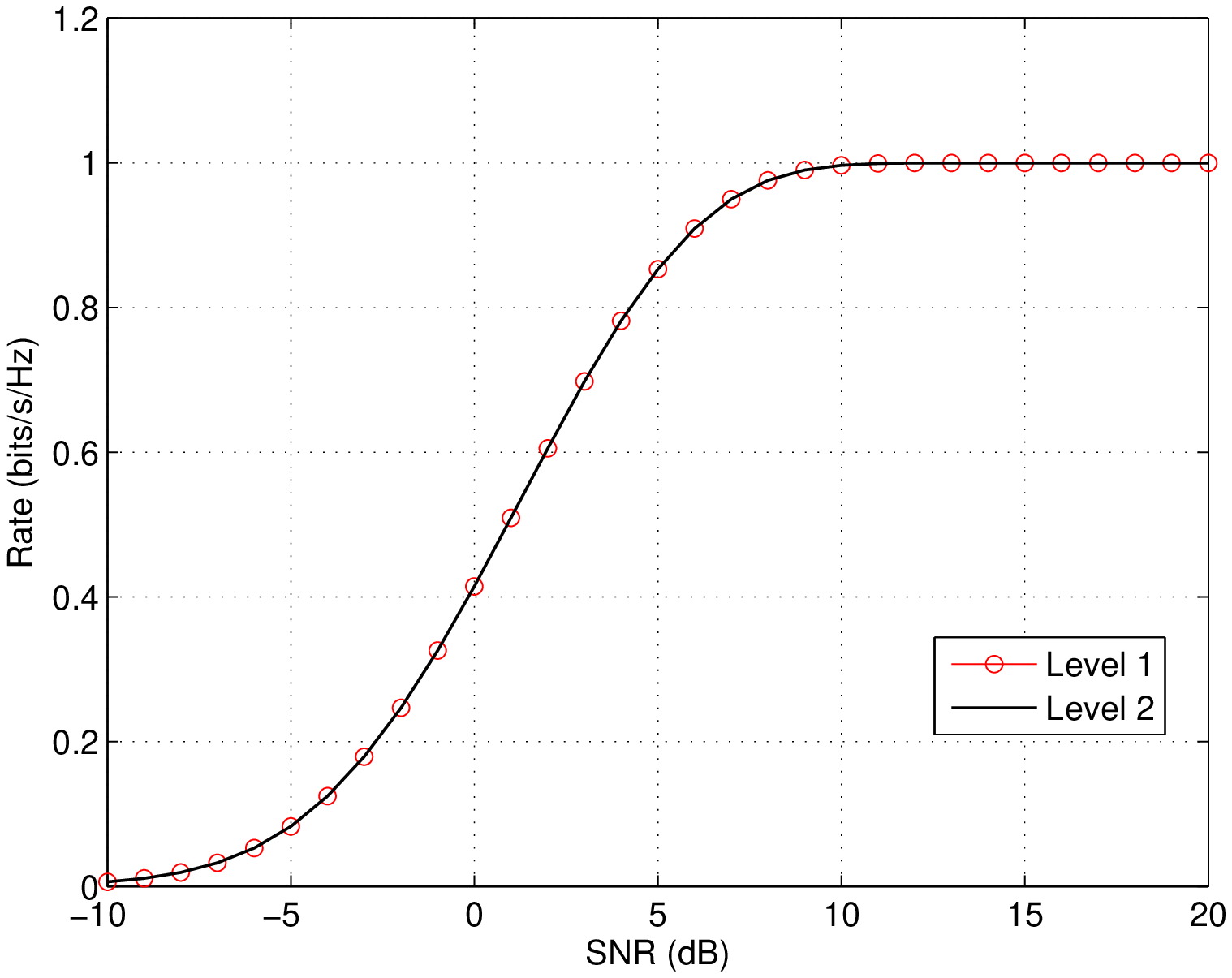}
\end{minipage}
\begin{minipage}{0.32\textwidth}
  \centering
	{\small \hspace{.3in} $8$-PSK }
  \includegraphics[width=1.12\hsize]{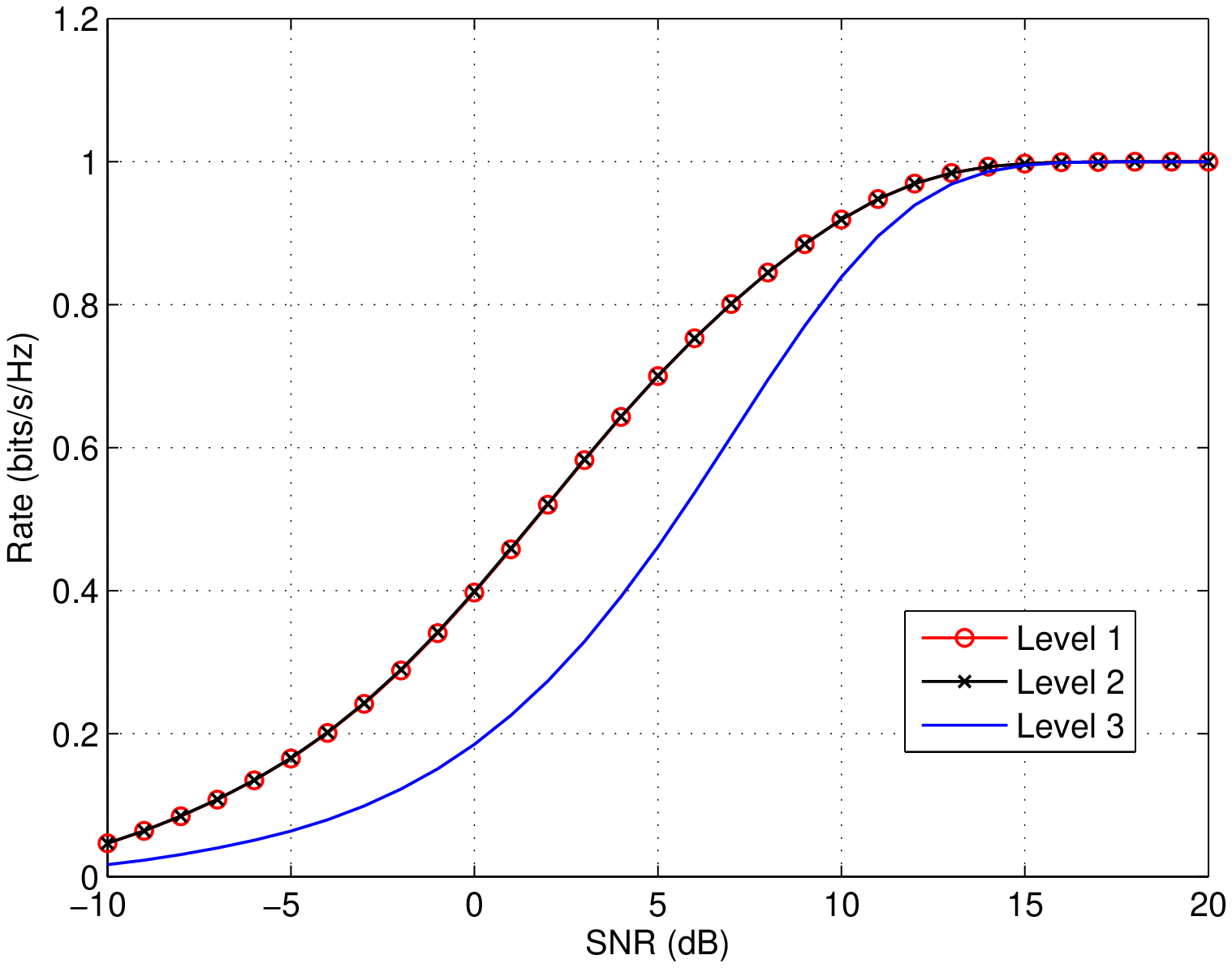}
\end{minipage}
\begin{minipage}{0.32\textwidth}
  \centering
	{\small \hspace{.3in} $16$-PSK }
  \includegraphics[width=1.12\hsize]{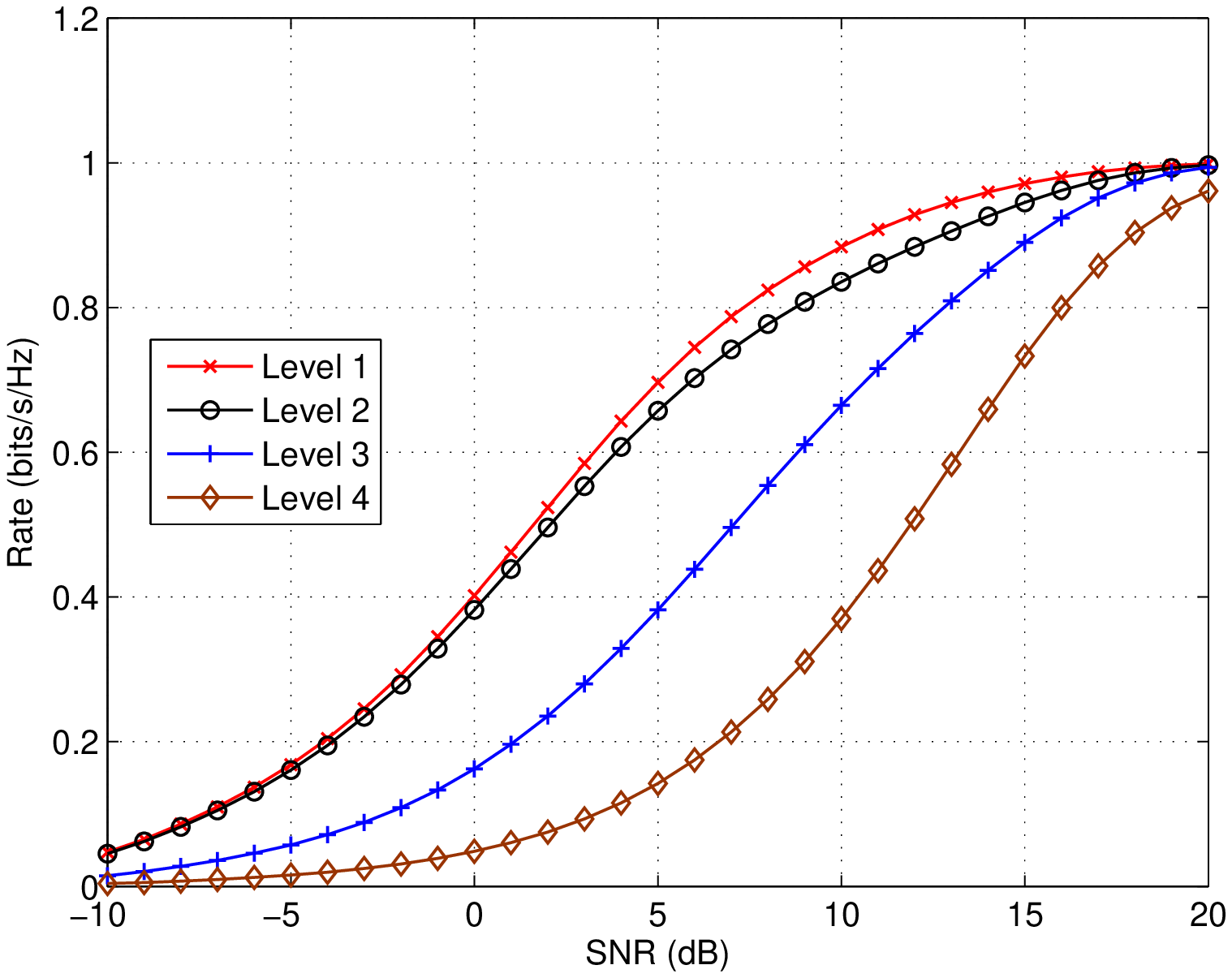}
\end{minipage}\\[0.2in]
\begin{minipage}{0.32\textwidth}
  \centering
	{\small \hspace{.3in} $8$-AMPM }
  \includegraphics[width=1.12\hsize]{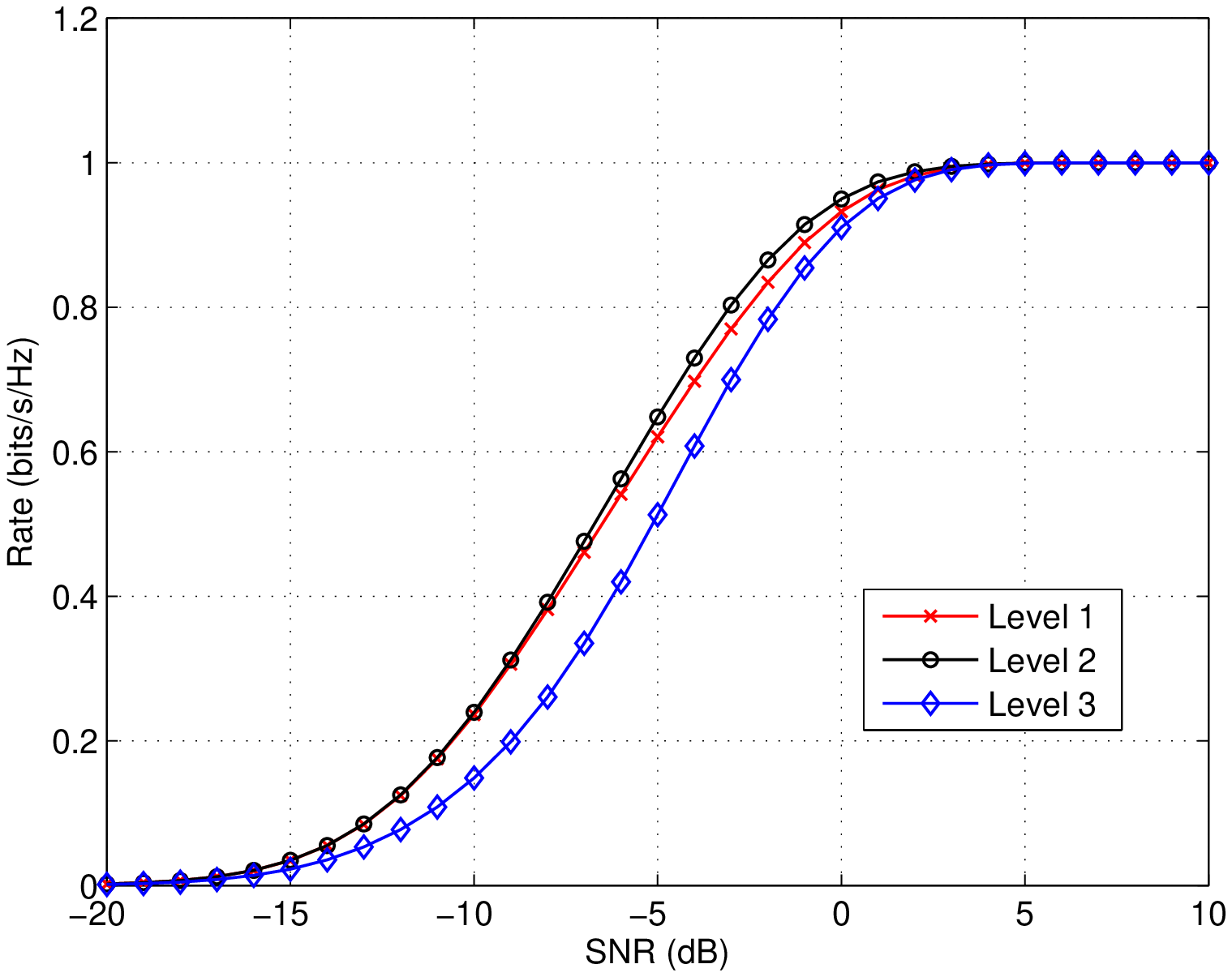}
\end{minipage}
\begin{minipage}{0.32\textwidth}
  \centering
	{\small \hspace{.3in} $16$-QAM }
  \includegraphics[width=1.12\hsize]{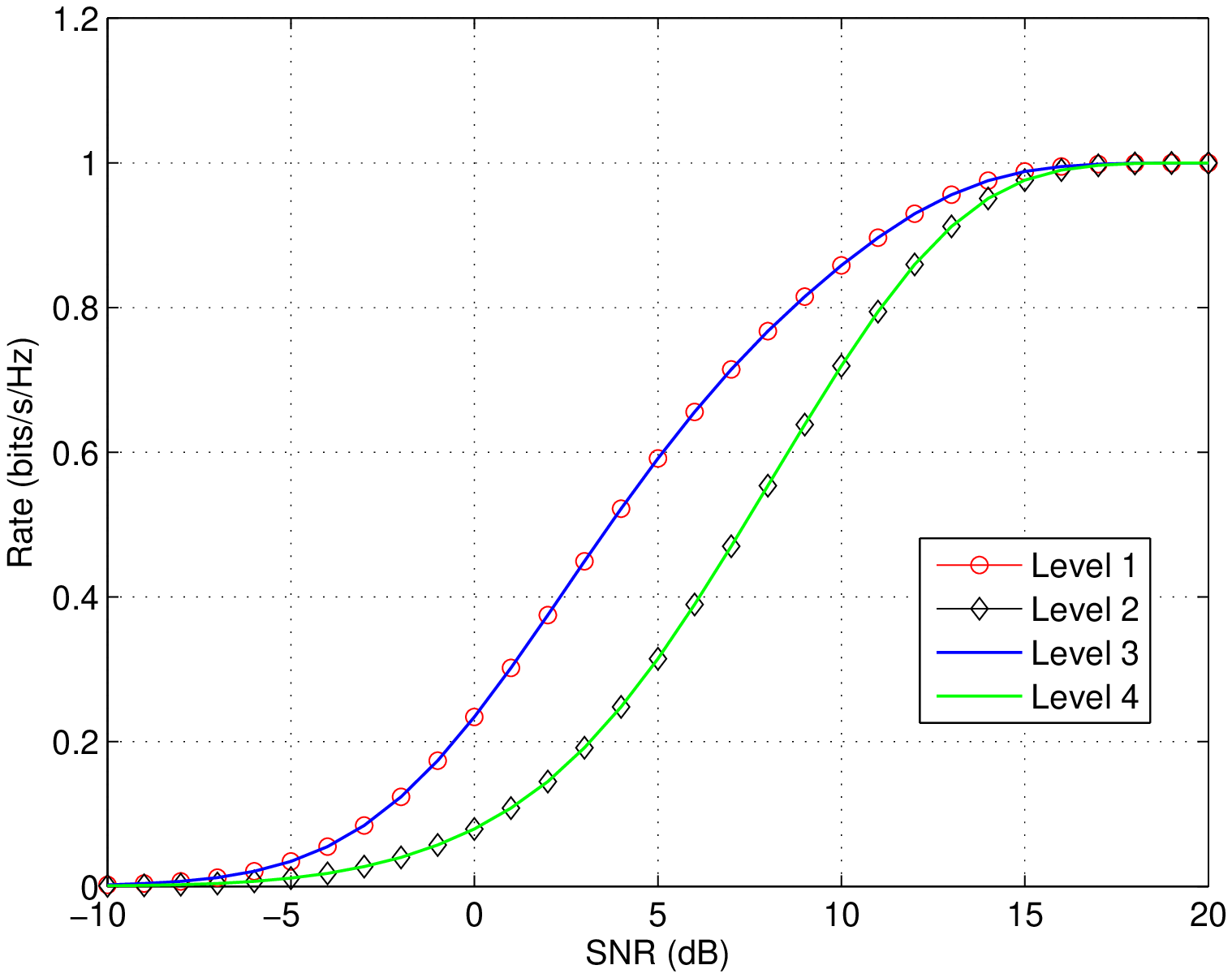}
\end{minipage}
\begin{minipage}{0.32\textwidth}
  \centering
	{\small \hspace{.3in} $32$-AMPM }
  \includegraphics[width=1.12\hsize]{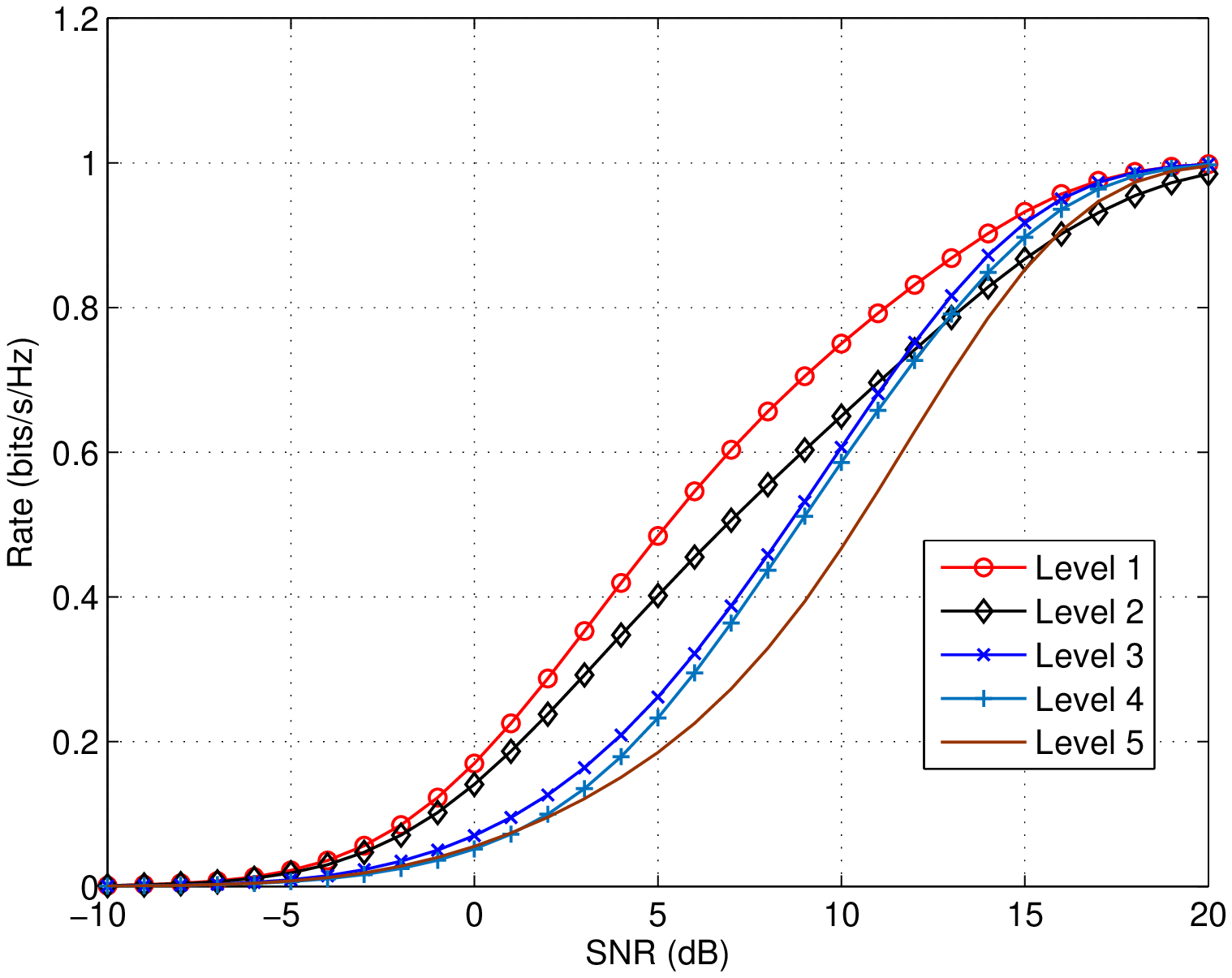}
\end{minipage}
\caption{Single-user MLC mutual information curves for a variety of  PAM, PSK and QAM-type constellations with natural mapping. MLC mutual information depends on decoding order, which in the case of these curves has been from the most to least significant bit of the modulation mapping. The broadcast users ``see'' such channels at respective operating points $\rho_1$ and $\rho_2$.}
\label{fig:Multilevel_p2p_capacity}
\end{figure*}

To summarize the developments so far: a pragmatic near-optimal rate allocation algorithm is being developed to allow the implementation of superposition coding in practical applications. So far, it was shown that the overall rate utility function as well as the constraints can be decomposed to level-wise utility and constraint functions that are minimally coupled (only through the shared parameter $\theta$). The main remaining computational aspect is the calculation of the level-wise constraints. 
Fortunately, the affine approximation allows us to characterize the level-wise constraints via their two end-points, and the insensitivity of each constraint to other levels' parameters allows us to obtain these end points from the single-user mutual information curves of multilevel modulations. 
We produce in Fig.\ref{fig:Multilevel_p2p_capacity} a series of such curves for PAM, PSK, and QAM type modulations. These curves may be pre-calculated and stored via lookup tables. Then the rate constraints at each level may be obtained by reading the values off these curves at the respective SNRs for the two channels.

\subsection{Exceptions to the Decoupling of Bit-level Rate Constraints}



\begin{figure}[h]
\centering
\includegraphics[scale=1.3]{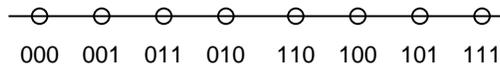}
\caption{$8$-PAM constellation with Gray-like mapping.}
\label{fig:Gray_like}
\end{figure}

The performance of the proposed rate allocation algorithm  is virtually indistinguishable from optimal for many practical cases including many familiar modulations under natural and Gray mapping. The excellent performance was explained via  the insensitivity of the bit-level rate constraints to the operating point in the other bit-levels. A key remaining question is: how prevalent is this insensitivity (decoupling) condition, and what is the performance penalty of the proposed algorithm when this condition does not hold? To our experience, counter-examples to this insensitivity condition are very rare and involve irregular mappings or constellations. As an example, we offer a Gray-like mapping for $8$-PAM as shown in Fig.~\ref{fig:Gray_like}. 


\begin{figure}
\centering
\includegraphics[width=\plotwidth]{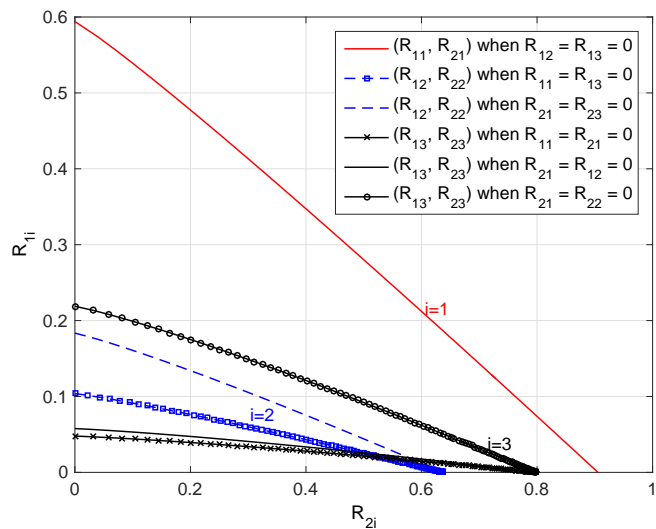}
\caption{Bit-level rate constraints for the Gray-like mapping of Fig~\ref{fig:Gray_like}.}
\label{fig:Gray_8PAM}
\end{figure}

\begin{figure}
\centering
\includegraphics[width=\plotwidth]{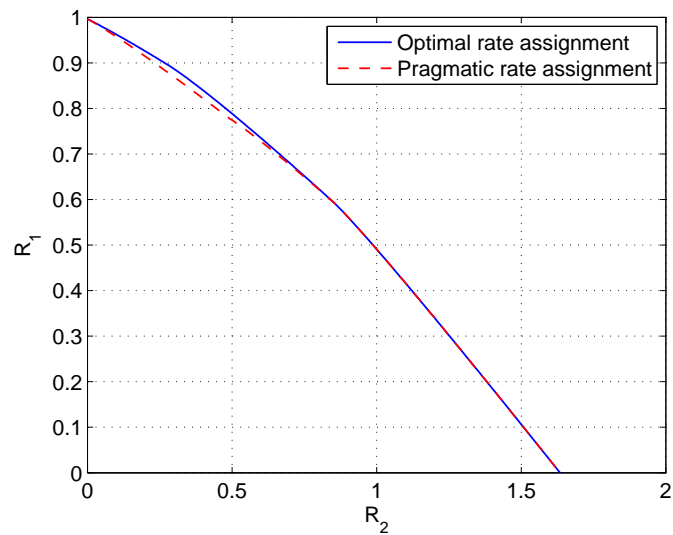}
\caption{Transmission rate using the general optimization versus the efficient optimization.}
\label{fig:general_vs_efficient}
\end{figure}

The sensitivity of the bit-level broadcast rate constraints for this modulation are demonstrated in Fig.~\ref{fig:Gray_8PAM}. It is observed that unlike the previous cases, the bit-level constraint of level $3$ is sensitive to the bit-level constraint in level $1$. This sensitivity manifests itself in a (slight) sub-optimality of the pragmatic rate allocation technique introduced in the previous subsection. Despite the apparent sensitivity, the resulting sub-optimality is slight and is demonstrated in Fig.~\ref{fig:general_vs_efficient}.

Of course an example does not make a general case, therefore in the interest of completeness, we outline in the remainder of this subsection a relaxation method can be used for allocating each level's rates to the two users, with no pre-determined constraints on the outcome of the rate allocation. Although it is our understanding that the previous subsection's pragmatic method should be sufficient for almost all practical cases.




The desired solution can be characterized in the form of two vectors $\boldsymbol{R}_1,\boldsymbol{R}_2$ whose components carry the components of the rates in individual levels dedicated to User~1 and User~2.

One way to think about solving this optimization problem is as follows. First, we assign all the rate to one of the receivers (without loss of generality receiver 2), such that
\begin{align}\nonumber
&\boldsymbol{R}_{1}=[0\dots 0]\\\nonumber
&\boldsymbol{R}_{2}=[C_{21}\dots C_{2m}] 
\end{align}
where  $C_{1i}$ and $C_{2i}$ denote the point-to-point capacity of level-$i$ for the weak receiver and the strong receiver respectively.



In order to move on the boundary of the capacity region so that receiver 1 is assigned a portion of the rate, each step should maximize the gain in $R_1$ while maintaining minimum loss to $R_2$.

This can be done by incrementing one of the entries of $\boldsymbol{R}_1$, i.e., increasing $R_{1i}$ for some $i$. However, the corresponding loss in $R_{2i}$ depends on the bit constraint of level $i$. Thus, it is reasonable to increment $R_1$ through level $i$ that provides maximum gain in $R_1$ given a fixed loss in $R_2$. The remaining task is finding a plausible choice of level $i$ as follows. First the bit-level constraint for each level $i$ and its slope denoted by $\bar{f}_i$ are calculated at the current rate assignment. Note that $\bar{f}_i$ represents the gain in $R_{1i}$ normalized to the loss in $R_{2i}$. The level $i^*$ that results in the maximum gain in $R_1$ satisfies
\begin{equation}\label{max_slope}
|\bar{f}_{i^*}|>|\bar{f}_j| \quad \forall_j.
\end{equation}

Therefore, moving close to the boundary of the capacity region can be realized by increasing $R_1$ through increasing $R_{1i^*}$ and fixing $R_{1j}$ $\forall j\neq i^*$ until either $R_{1i^*}$reaches its maximum value $C_{1i^*}$ or the inequality \eqref{max_slope} is violated. In either case, the same procedure is then repeated until the desired rate pair is achieved.

\subsection{Multilevel BICM construction}
\begin{figure}
\centering
\includegraphics{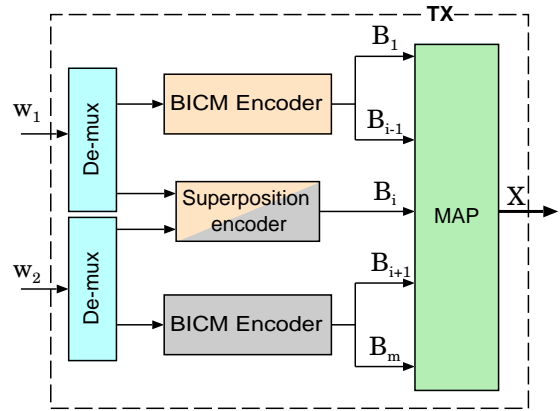}
\caption{Hybrid MLC-BICM superposition}
\label{fig:MLC_BICM_BC}
\end{figure}

BICM is a close relative of MLC in the point-to-point channel, where the bits from multiple levels are encoded using not only the same code rate, but together as one code word. In our proposed multilevel superposition coding with the efficient structure shown in Fig.~\ref{fig:MLC_BC_Efficient}, there are $m$ encoders: some of them carry information for the weak receiver, some of them carry information for the strong receiver and at most one encoder that carries information for both receivers. We propose to combine all the encoders that carry information for a certain receiver in one BICM encoder as shown in Fig.~\ref{fig:MLC_BICM_BC}. This way of transmission reduces the number of encoders significantly especially for big constellations. For example, for a $64$-QAM constellation, the multilevel coding structure will require at least six encoders and by combining all the encoders that send to the same receiver into one BICM encoder, the number of encoders can be reduced to at most three encoders but with longer block length. We call this technique the hybrid technique since it uses multilevel coding in the sense of encoding the information independently and BICM encoder to encode the information that belong to the same receiver.

\begin{figure}
\centering
\includegraphics[width=\plotwidth]{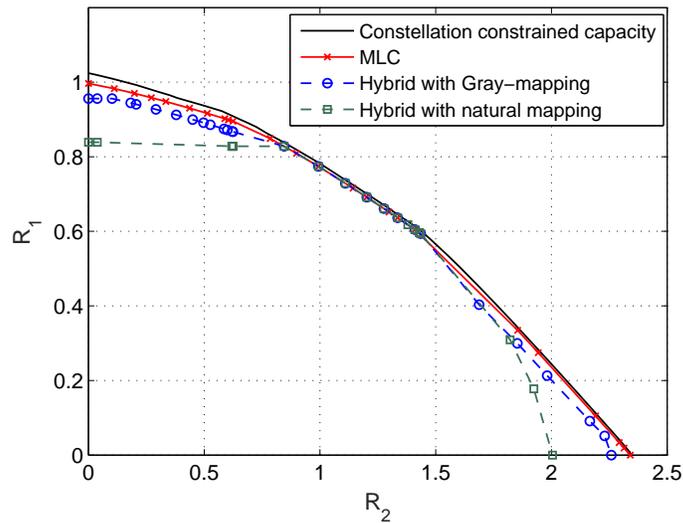}
\caption{MLC and hybrid superposition achievable rates under $8$-PAM, $\rho_1=5dB$, $\rho_2=15dB$.}
\label{fig:MLC_vs_BICM}
\end{figure}

The rate of the BICM encoder and the serial to parallel conversion depends on the number of levels that the encoder feeds. The rate achieved by the hybrid transmission is shown in Fig.~\ref{fig:MLC_vs_BICM} for Gray and natural mappings. The achievable rate region of the hybrid transmission is in general smaller than the achievable rate region of the multilevel coding scheme since BICM is not capacity achieving. The maximum loss in rate is the point-to-point transmission since the encoding becomes completely point-to-point BICM encoding; however, when the rates of the weak and the strong receivers are not equal to zero, the transmission becomes closer to the multilevel superposition transmission. For example for the $8$-PAM constellation, there is a stage in which the MLC and Hybrid schemes will be the same. This is the point when the level that carries information for both receivers is the middle level.

\section{Simulations}

Because the broadcast channel involves simultaneously two rates and two SNRs, error plots are generated for the broadcast channel by applying slight modifications to the standard methods used for plotting errors in point-to-point coding literature. For broadcasting the relative quality of the channels, indicated by the noise variances, remains fixed in the simulations, while the transmit power is allowed to increase. The rate of the two codes is chosen according to a rate pair on the boundary of the capacity region.  In each plot, the value of the transmit power corresponding to the capacity rate pair is clearly marked, a point that is the counterpart to the ``capacity threshold'' in the single-user error curves seen in the coding literature. A comparison between this point and the waterfall region of the error curves is an indicator of how far from optimality is the system operating.

\begin{figure}
\centering
\includegraphics[width=\plotwidth]{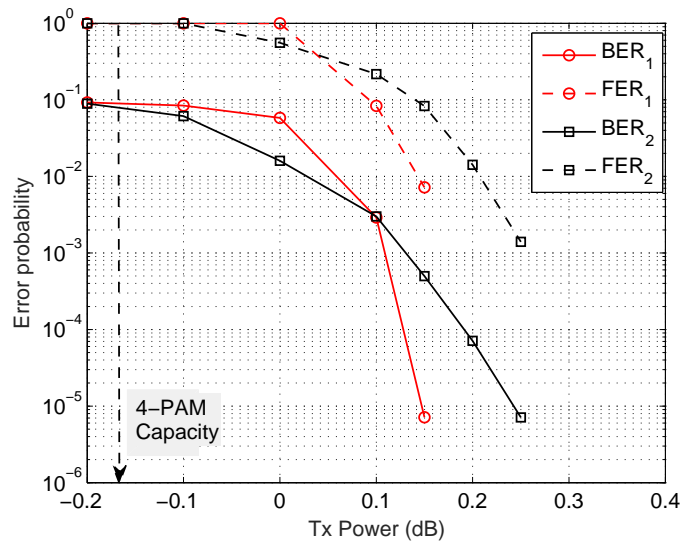}
\caption{Performance of Multilevel superposition for $4$-PAM constellation where $\sigma_1^2=.48$, $\sigma_2^2=.13$}
\label{fig:BER_4PAM}
\end{figure}

The DVB-S2 LDPC codes are used as component codes for each of the levels to examine the performance of the proposed MLC and the hybrid (MLC-BICM) transmissions. The block length of the codes is $n=64$k. Fig.~\ref{fig:BER_4PAM} shows the performance of $4$-PAM MLC superposition for rates $(R_2=0.5, R_1=0.6)$ with natural mapping. The information of the weak receiver is sent over level-1 and the information of the strong receiver is sent over level-2. This is considered an extreme case where each level is assigned to either the weak or the strong receiver. The bit error rate (BER) and frame error rate (FER) for each receiver are shown. The gap to capacity is approximately $0.5$-dB at $10^{-5}$ FER, which is the similar to the gap to capacity of the DVB code in the point-to-point channel, thus suggesting that the FER gap is mostly due to the limitations of the code as opposed to the MLC.

\begin{figure}
\centering
\includegraphics[width=\plotwidth]{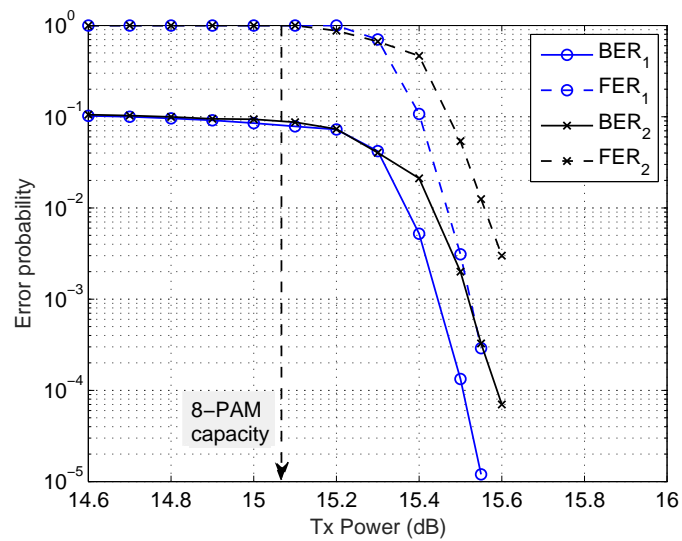}
\caption{Performance of Multilevel superposition for $8$-PAM constellation where $\sigma_1^2=8.5$, $\sigma_2^2=1$}
\label{fig:BER_8PAM}
\end{figure}

Fig.~\ref{fig:BER_8PAM} shows the performance of $8$-PAM constellation where one bit level is shared between the weak and the strong receiver. The rates assigned are $R_1=0.6$ and $R_2=1.4$. Level-1 carries information only for the weak receiver, level-2 is shared, and level-3 carries information only for the strong receiver. In the shared level, the weak and the strong receivers messages are encoded independently using the DVB-S2 LDPC codes and combined after setting some bits of the strong receiver codeword to zeros as described in Section~\ref{sec:Bit-wise}.

\begin{figure}
\centering
\includegraphics[width=\plotwidth]{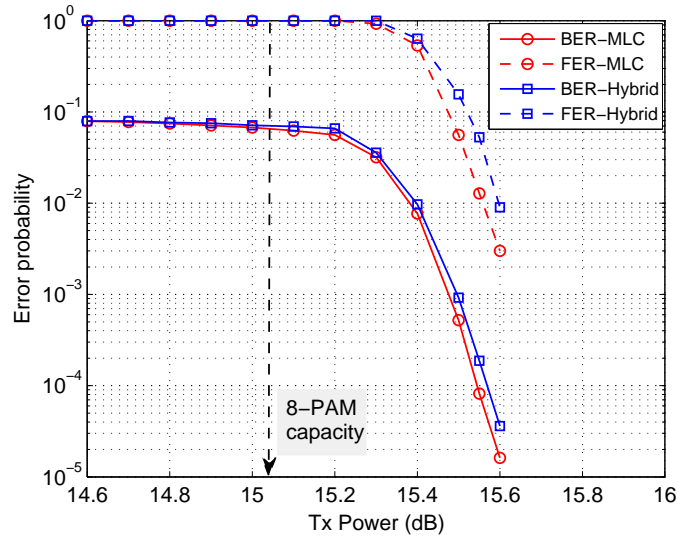}
\caption{Performance of the hybrid MLC-BICM scheme for $8$-PAM constellation where $\sigma_1^2=8.5$, $\sigma_2^2=1$}
\label{fig:BER_MLC_BICM}
\end{figure}

Fig.~\ref{fig:BER_MLC_BICM} shows the BER and FER of the proposed hybrid MLC-BICM (Fig.~\ref{fig:MLC_BICM_BC}) transmission compared with the MLC transmission (Fig.~\ref{fig:MLC_BC_Efficient}) for an $8$-PAM constellation with Gray mapping. Level-1 carries information for the weak receiver and the other two levels carry information for the strong receiver. The rates are $R_1=0.5$ and $R_2=1.5$. In the hybrid transmission, a BICM encoder is used with double the length of the one used in level-1 and the output of the BICM encoder is partitioned into two streams and fed to the two least significant bits. Simulation  show that the hybrid scheme has a performance very close to that of MLC.

\begin{figure}
\centering
\includegraphics[width=\plotwidth]{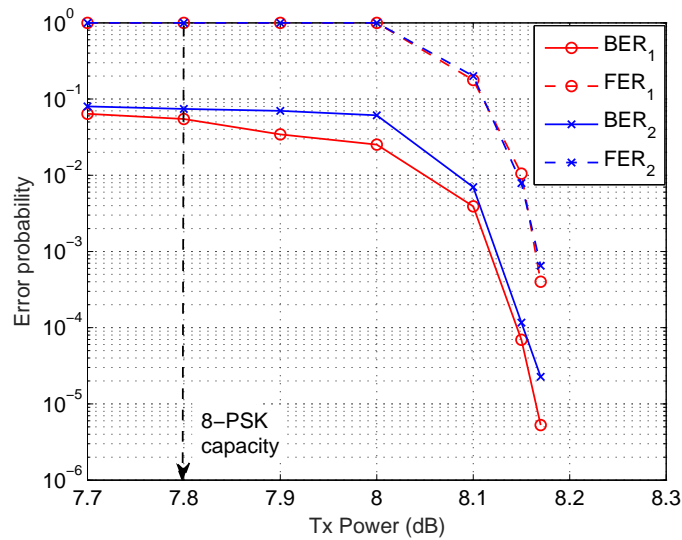}
\caption{Performance of the MLC proposed transmission for $8$-PSK constellation where $\sigma_1^2=2.2$, $\sigma_2^2=1$}
\label{fig:BER_8PSK}
\end{figure}

\begin{figure}
\centering
\includegraphics[width=\plotwidth]{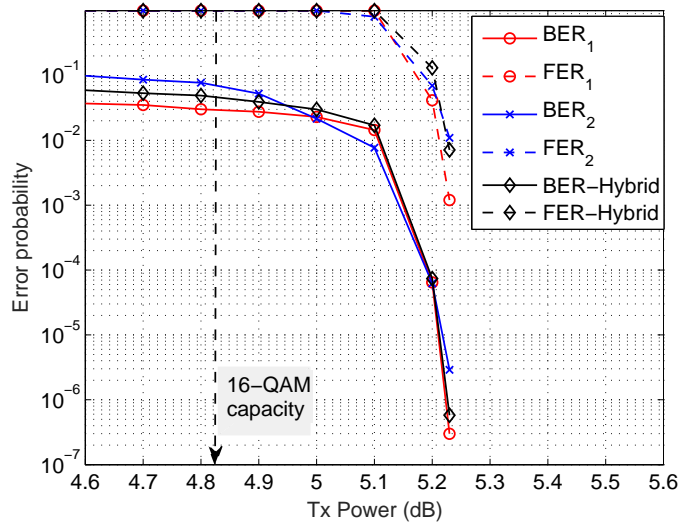}
\caption{Performance of the MLC proposed transmission and the Hybrid MLC-BICM transmission for $16$-QAM constellation where $\sigma_1^2=.64$, $\sigma_2^2=.18$}
\label{fig:BER_16QAM}
\end{figure}

Fig.~\ref{fig:BER_8PSK} shows the error performance of  $8$-PSK constellation with natural mapping where level-1 carries information for the weak receiver, level-3 carries information for the strong receiver and level-2 carries information for both receivers. The rates are $R_1=0.4$ and $R_2=1.6$. The gap to capacity is around $0.5$-dB at bit error probability of $10^{-5}$.

Fig.~\ref{fig:BER_16QAM} shows the performance of $16$-QAM constellation with natural labeling where level-1 carries information for the weak receiver, level-2 for both receivers, and levels 3 and 4 carry information for the strong receiver. The rates are $R_1=1.2$ and $R_2=1.8$ and noise variances at the two receivers are $\sigma_1^2=.64$ and $\sigma_2^2=.18$. The simulations show that the proposed scheme has a gap of around $0.4$-dB from the constellation constrained capacity at bit error probability of $10^{-5}$. The figure also shows the performance of the Hybrid MLC-BICM transmission where the two encoders of the two least significant bits are combined in one BICM encoder while using Gray mapping.

\section{conclusion}

This paper studied coded modulation for the AWGN broadcast channel. multilevel coding (MLC) and bit-interleaved coded modulation (BICM) are explored under channel-input modulation constraints. It was shown that the assignment of receivers information to distinct inputs to the mapper does not approach the capacity uniformly. A bit-wise multilevel superposition transmission is proposed. Furthermore, a hybrid MLC-BICM with lower complexity is proposed. The achievable rate region of the proposed transmission is very close to the boundary of the constellation constrained capacity of the broadcast channel. Simulation results showed an excellent performance using good point-to-point codes.

\appendices

\section{Degradedness of bit channels}

Consider the following Markov process due to the degradedness of the channel
\[
U \rightarrow X \rightarrow Y_2 \rightarrow Y_1
\]

$U$ has a multi-digit characterization $[C_1, \dots , C_m]$.

for a specific value of $C^{i-1}=c^{i-1}$, due to the degradedness of the channel we have
\[
I(C_i;Y_1|C^{i-1}=c^{i-1}) \leq I(C_i;Y_2|C^{i-1}=c^{i-1})
\]
The mutual information $I(C_i;Y_1|C^{i-1})$ and $I(C_i;Y_2|C^{i-1})$ are
\begin{align}\label{deg}
I(C_i;Y_1|C^{i-1})=E_{C^{i-1}}[I(C_i;Y_1|C^{i-1}=c^{i-1})]\\
I(C_i;Y_2|C^{i-1})=E_{C^{i-1}}[I(C_i;Y_2|C^{i-1}=c^{i-1})]
\end{align}
where $E[.]$ is the expectation operation. The expectation operation is a convex combination for all the values that $C^{i-1}$ can take. Since the inequality \eqref{deg} holds for any value of $C^{i-1}$ then it holds for any convex combination of the values of $C^{i-1}$, therefore:
\[
I(C_i;Y_1|C^{i-1})\leq I(C_i;Y_2|C^{i-1})
\]

\section{Multilevel Decomposition of the Outer Code}
\label{Appendix:OuterCodeDecomposition}

Consider the auxiliary random variable $U$ representing the message to the weak user. To achieve capacity, the outer code is drawn i.i.d. according to $p_U(u)$. In the following we assume the cardinality $|U|=M$. The objective is to produce multilevel codes whose empirical distribution approaches $p_U(u)$. We now consider an $m$-dimensional  binary vector $V$ whose components are i.i.d. Bernoulli-$\frac{1}{2}$. Equivalently, $V$ can be considered a random variable uniformly distributed over an alphabet size of $2^m$. This is the random variable generating the $m$-level multilevel code. Consider the design of a mapping $U'=f(V)$ so that the random variable $U'$, in distribution, is close to the capacity-maximizing $U$. We start with:
\[
p_U(u) = [ p_1 \cdots,p_M]
\]
Rounding down each of the probabilities to a multiple of $2^{-m}$ via $Q(p_i)\triangleq2^{-m} \lfloor 2^m p_i\rfloor$, and distributing the remaining probability $1-\sum_i Q(p_i)$ over the first $K \triangleq 2^m(1-\sum_i Q(p_i))$ components, we arrive at the following probability distribution for $U'$:
\[
p_{U'}(i) = \begin{cases}
Q(p_i)+2^{-m}   & i\leq K \\
Q(p_i) & i> K
\end{cases}
\]
Defining $k_i \triangleq 2^{m} p_{U'}(i)$, the function $f(\cdot)$ given below maps the multilevel binary generator variable $V$ to the (approximate) capacity achieving distribution $U'$:
\[
f(j)=\begin{cases}
1 & 1\le j < k_1\\
2 & k_1 \le j < k_1+k_2\\
\cdots\\
M & k_1+\cdots+k_{M-1} \le j < k_1 +\cdots +k_M
\end{cases}
\]
In the following, we assume that none of the entries of $p_U$ are zero, and also that $m$ is large enough so that none of the entries of $p_{U'}$ are zero. A sufficient condition is $m > - \log_2 \min_i p_U(i)$.

Now, it is straightforward to bound the divergence between $p_U$ and $p_{U'}$:
\begin{align*}
D(p_U||p_{U'}) &= \sum_i P_U(i) \log \frac{P_U(i)}{P_{U'}(i)} \\
&\le \sum_i P_U(i) \log \frac{P_{U'}(i)+2^{-m}}{P_{U'}(i)} \\
&\stackrel{(a)}{\le} \sum_i P_U(i) \frac{2^{-m}}{P_{U'}(i)}\\
&\stackrel{(b)}{\le} M 2^{-m+1}
\end{align*}
where $(a)$ follows from $\log(1+x)\le x$ and $(b)$ follows from $\frac{p_U(i)}{p_{U'}(i)} \le \frac{p_U(i)}{Q(p_{U}(i))} \le 2$.

Therefore, it follows that for a fixed $M$, by increasing the number of levels $m$ one can very quickly get close to the capacity optimizing distribution.

\bibliographystyle{IEEEtran}
\bibliography{IEEEabrv,Attia}

\end{document}